\documentclass[a4paper,11pt]{article}
\pdfoutput=1 

\usepackage{jheppub} 

\usepackage[T1]{fontenc} 

\usepackage[dvipsnames]{xcolor}

\usepackage{enumitem}

\usepackage{mathrsfs}

\newcommand{\ket}[1]{\left|#1\right\rangle}

\newcommand{\bei}{\begin{itemize}}
\newcommand{\eei}{\end{itemize}}
\newcommand{\bee}{\begin{enumerate}}
\newcommand{\eee}{\end{enumerate}}
\newcommand{\beeL}{\begin{enumerate}[label=(\Alph*)]}
\newcommand{\beel}{\begin{enumerate}[label=(\alph*)]}
\newcommand{\beeR}{\begin{enumerate}[label=(\Roman*)]}
\newcommand{\beer}{\begin{enumerate}[label=(\roman*)]}
\newcommand{\beeLd}{\begin{enumerate}[label=\Alph*.]}
\newcommand{\beeld}{\begin{enumerate}[label=\alph*.]}
\newcommand{\beeRd}{\begin{enumerate}[label=\Roman*.]}
\newcommand{\beerd}{\begin{enumerate}[label=\roman*.]}

\def\a {{\alpha}}

\def\bbf{{a}}

\usepackage{tikz}
\usetikzlibrary{math}
\usetikzlibrary{arrows,shapes,positioning}
\usetikzlibrary{decorations.markings}
\tikzstyle arrowstyle=[scale=1]
\tikzstyle directed=[postaction={decorate,decoration={markings,
    mark=at position .65 with {\arrow[arrowstyle]{stealth}}}}]
\tikzstyle reverse directed=[postaction={decorate,decoration={markings,
    mark=at position .65 with {\arrowreversed[arrowstyle]{stealth};}}}]

\newlength{\mywidth}
\newcommand{\bal}{\begin{equation}\begin{aligned}}
\newcommand{\eal}{\end{aligned}\end{equation}}

\renewcommand{\L}{{\scriptscriptstyle\text{L}}}
\newcommand{\R}{{\scriptscriptstyle\text{R}}}

\newcommand{\de}{\text{d}}

\title{\boldmath 
Thermodynamics of integrable $\mathcal{N}=2$ theories, squared
}

\author[1]{Xavier Kervyn,} 
\author[2,3,4]{Davide Polvara,}
\author[3,4]{Alessandro Sfondrini}

\affiliation[1]{Arnold Sommerfeld Center for Theoretical Physics, Ludwig-Maximilians-Universität, 80333 Munich, Germany.}
\affiliation[2]{II. Institut f\"ur Theoretische Physik,  Universit\"at Hamburg, Luruper Chaussee 149, 22761 Hamburg, Germany.}
\affiliation[3]{Dipartimento di Fisica e Astronomia, Universit\`a degli Studi di Padova, via Marzolo 8, 35131 Padova, Italy.}
\affiliation[4]{Istituto Nazionale di Fisica Nucleare, Sezione di Padova, via Marzolo 8, 35131 Padova, Italy.}

\emailAdd{xavier.kervyn@campus.lmu.de}
\emailAdd{davide.polvara@gmail.com}
\emailAdd{alessandro.sfondrini@unipd.it}

\abstract{
In arXiv:2306.17553 a new supersymmetric integrable QFT was constructed from the relativistic limit of the worldsheet theory of AdS$_3 \times$ S$^3\times $T$^4$ superstrings with mixed Ramond-Ramond and Neveu-Schwarz-Neveu-Schwarz flux.
The model is closely reminiscent of one previously considered by Fendley and Intriligator, though it enjoys twice as many supersymmetries.
In this paper we study its finite-volume and finite-temperature properties.
We formulate the string hypothesis for the Bethe--Yang equations, write down the thermodynamic Bethe ansatz equations, and simplify them in the form of a Y-system.
We obtain three decoupled sectors associated with massive, massless-chiral and massless-antichiral particles, for which we compute the UV central charge. We illustrate the similarities and differences between this model and the one of Fendley and Intriligator.
}

\begin{document} \begin{flushright}\small{ZMP-HH/25-4}\end{flushright}
 
\maketitle
\flushbottom

\newpage
\section{Introduction}
\label{sec:introduction}

Superstrings propagating in AdS$_3 \, \times$ S$^3\times \, $T$^4$  have been an important instance of the holographic correspondence since its discovery~\cite{Maldacena:1997re}.
This setup is richer than the better-studied AdS$_5 \times $ S$^5$ since it has half of its supersymmetry. As a result, the supergravity background with AdS$_3 \, \times$ S$^3\times \, $T$^4$ may be supported by arbitray combinations of Neveu-Schwarz-Neveu-Schwarz (NSNS) and Ramond-Ramond (RR) fluxes. 
In perturbative string theory, the amount of NSNS flux is quantised in units of~$k$, while the strength of the RR forms is not, and is proportional to~$h\geq0$. The string tension $T$ is
\begin{equation}
    T=\sqrt{\frac{k^2}{4\pi}+h^2}\,,
\end{equation}
and one can treat $k=0,1,2,\dots$ and $h\geq0$ as parameters of the classical string non-linear sigma model (NLSM)\footnote{
There are further parameters but up to TsT transformations~\cite{Lunin:2005jy,Frolov:2005dj,Alday:2005ww} the parameters $k,h$ are sufficient to capture the structure of the spectrum; see~\cite{OhlssonSax:2018hgc} for a discussion of the parameters of the model and~\cite{Seibold:2024qkh} for a review of its integrable deformations.}. In~\cite{Cagnazzo:2012se} it was shown that this family of NLSM is classically integrable for all choices of fluxes, allowing in principle to apply the integrability machinery to extract the string spectrum exactly, as it was done for AdS$_5 \times $ S$^5$ (see~\cite{Beisert:2010jr,Arutyunov:2009ga} and references therein).
At the special points where the background does not involve any RR flux, the worldsheet theory is a supersymmetric Wess-Zumino-Witten model and the spectrum can be found in a closed form~\cite{Maldacena:2000hw}).\footnote{See also~\cite{Dei:2018jyj} for the derivation of the pure-NSNS spectrum through integrability.}
In that case the spectrum is much simpler and more degenerate than the one of e.g.~AdS$_5\times$S$^5$. Conversely, in the presence of RR flux the spectrum is believed to be much more general, with no additional degeneracy other than the ones due to the manifest symmetries of the model.
Currently, it appears that only worldsheet integrability may provide a way to obtain the spectrum in this more general setup.%
\footnote{See however~\cite{Cho:2018nfn} for a discussion of the worldsheet-CFT approach in the presence of RR flux.}
In this way, much progress has been recently achieved in the pure-RR case, where a complete proposal for the worldsheet S-matrix (comprising the dressing factors) now exists~\cite{Frolov:2021fmj,Frolov:2021zyc}, TBA equations have been derived~\cite{Frolov:2021bwp} and a proposal for a Quantum Spectral Curve has also been advanced~\cite{Ekhammar:2021pys,Cavaglia:2021eqr,Cavaglia:2022xld}. The mixed flux case (when both the RR and NSNS fluxed are turned on) is harder, due to the highly complicated dispersion relation of the worldsheet particles~\cite{Hoare:2013lja,Lloyd:2014bsa}. In this case only the matrix part of the S-matrix is known~\cite{Lloyd:2014bsa} while the dressing phases have been recently bootstrapped only for the interaction of massive modes~\cite{Frolov:2024pkz,Frolov:2025uwz} (partial results were also found in~\cite{OhlssonSax:2023qrk}). 

In this paper we consider a model arising in the relativistic limit of the mixed-flux string, for which the bootstrap program was completed by Frolov, Polvara and Sfondrini (FPS)~\cite{Frolov:2023lwd}. Remarkably, this construction included the S~matrix ``dressing factors'', owing to the fact that in the relativistic limit it is substantially easier to solve the crossing equations than in the original non-relativistic string model. 
The FPS model was obtained by expanding the kinematics of the full model in a low-energy and small-$h$ regime. Among other things, it provides a check for the S-matrix bootstrap program of the full theory. 
Moreover, it is a self-consistent supersymmetric integrable theory, interesting in its own right. Indeed, the study of form factors for this model was recently initiated in~\cite{Torrielli:2023uam}.
The FPS model is reminiscent of a model studied by Fendley and Intriligator (FI) in~\cite{Fendley:1991ve,Fendley:1992dm}, though there are some crucial differences: in particular, the FPS model enjoys twice as many supersymmetries than the FI one. In a sense, this model is a ``square'' of the one of~\cite{Fendley:1991ve,Fendley:1992dm} --- its S~matrix can be constructed from tensoring two copies of the FI S~matrix.
In this paper we will derive the TBA equations and Y system for the FPS model and use them to extract the UV central charge of the model at all values of~$k$, comparing the result with the ones of FI. The comparison is quite non-trivial: for instance, the UV central charge of this model \textit{is not} twice that of FI, as we will see.

The paper is structured as follows. In section~\ref{sec:rel_BY} we present the Bethe--Yang equations of the worldsheet mixed-flux theory, which can be trivially derived from the ones presented in~\cite{Seibold:2022mgg} by introducing the parameter $k$ in the left and right Zhukovsky variables. Taking the relativistic limit of these equations and dressing them with the known proposals for the dressing factors~\cite{Frolov:2023lwd} we obtain a complete set of Bethe--Yang equations for the relativistic mixed-flux model. Starting from these equations, in section~\ref{sec:TBA} we derive the TBA equations of the model and associated Y system, which in section~\ref{sec:central_charge} we use to derive the ultraviolet central charge of the theory. We conclude in section~\ref{sec:conclusions}. Additionally we provide some appendices where we give some more detail on the computations, such as a direct derivation of the Bethe--Yang equations in appendix~\ref{app:first_appendix}, a discussion of the Bethe strings of the model (which are just its bound states) in appendix~\ref{app:BS}, a review of the derivation of the TBA equations and Dilogarithm trick for the extraction of the central charge in appendices~\ref{app:derivation_TBA} and~\ref{sec:app_Rogers}, respectively, and the derivation of the Y system in appendix~\ref{app:simplification_TBA}.

\section{Relativistic limit of the Bethe--Yang equations}
\label{sec:rel_BY}
In this section, we derive the Bethe--Yang equations for the scattering of excitations of superstrings propagating in the background AdS$_3 \, \times$ S$^3\times \, $T$^4$ with mixed flux in the relativistic limit considered in~\cite{Frolov:2023lwd}.  We will recover these equations from the ones of the full model written in~\cite{Seibold:2022mgg}. The equations can also be derived from scratch for the relativistic model as discussed in~\ref{app:first_appendix}.

An important remark is that to obtain the Bethe--Yang equations in the relativistic limit we start form the \textit{string} Bethe--Yang equations of the full model, as opposed to the \textit{mirror} Bethe--Yang equations. Indeed, the limit which we consider is inherently defined in the string kinematics.
The distinction between string and mirror theory is crucial in the original model, because it is non-relativistic~\cite{Arutyunov:2007tc}. After the limit such a distinction disappears (up to picking the boundary conditions on the circle, which depend on whether we consider a finite-volume or thermal partition function). This suggests that taking the relativistic limit and going to the mirror theory are non-commuting operations --- something to be kept in mind for the physical interpretation of our results.

\subsection{Bethe--Yang equations}
\label{sec:Bethe_Eq_before_limit}

We start by writing the Bethe--Yang equations for the scattering of fundamental particles in the full model, before taking any type of limit. These equations are obtained straightforwardly by generalising the results of~\cite{Frolov:2021fmj,Seibold:2022mgg} to the mixed flux case, for which we just need to introduce left and right Zhukovsky variables
\begin{equation}
\label{eq:zhukovsky_massive}
\begin{aligned}
    x^\pm_{\L}(p)&\equiv&
    \frac{\bigl( 1+\frac{k}{2 \pi} p \bigl) + \sqrt{\bigl( 1+\frac{k}{2 \pi} p \bigl)^2+4 h^2\sin^2(p/2)}}{2h\sin(p/2)} e^{\pm i \frac{p}{2}} \,,\\
    x^\pm_{\R}(p)&\equiv&
    \frac{\bigl( 1-\frac{k}{2 \pi} p \bigl) + \sqrt{\bigl( 1-\frac{k}{2 \pi} p \bigl)^2+4 h^2\sin^2(p/2)}}{2h\sin(p/2)} e^{\pm i \frac{p}{2}} \,,
\end{aligned}
\end{equation}
and Zhukovsky variables for massless particles
\begin{equation}
\label{eq:zhukovsky_massless}
    x^\pm(p) \equiv
    \frac{\frac{k}{2 \pi} p + \sqrt{\frac{k^2}{4 \pi^2} p^2+4 h^2\sin^2(p/2)}}{2h\sin(p/2)} e^{\pm i \frac{p}{2}} \,.
\end{equation}
Following the notation of~\cite{Seibold:2022mgg} we label the S-matrix elements associated with the scattering of the highest-weight states as follows
\begin{equation}
\label{eq:massivenorm}
    \begin{aligned}
    \mathbf{S}\,\big|Y_{p_1}Y_{p_2}\big\rangle&=
    S^{11}_{su}(p_1, p_2)\,
    \big|Y_{p_2}Y_{p_1}\big\rangle,\\
    \mathbf{S}\,\big|Y_{p_1}\bar{Z}_{p_2}\big\rangle&=
     \Tilde  S^{11}_{su}(p_1, p_2)\,
    \big|\bar{Z}_{p_2}Y_{p_1}\big\rangle,\\
    \mathbf{S}\,\big|\bar{Z}_{p_1}Y_{p_2}\big\rangle&=
 \Tilde  S^{11}_{sl}(p_1, p_2)\,
    \big|Y_{p_2}\bar{Z}_{p_1}\big\rangle,\\
    \mathbf{S}\,\big|\bar{Z}_{p_1}\bar{Z}_{p_2}\big\rangle&=
   S^{11}_{sl}(p_1, p_2)\,
    \big|\bar{Z}_{p_2}\bar{Z}_{p_1}\big\rangle .
    \end{aligned}
    \qquad \text{(massive-massive)}
\end{equation}

\begin{equation}
\label{eq:massivemasslessnorm}
    \begin{aligned}
    &\mathbf{S}\,\big|Y_{p_1}\chi^{\dot{\alpha}}_{p_2}\big\rangle= 
    S^{10}(p_1, p_2)\,
    \big|\chi^{\dot{\alpha}}_{p_2}Y_{p_1}\big\rangle,\\
 &\mathbf{S}\,\big|\bar{Z}_{p_1}\chi^{\dot{\alpha}}_{p_2}\big\rangle=
 \bar{S}^{10}(p_1, p_2)\,
    \big| \chi^{\dot{\alpha}}_{p_2} \bar{Z}_{p_1} \big\rangle,\\
    &\mathbf{S}\,\big|\chi^{\dot{\alpha}}_{p_1} Y_{p_2} \big\rangle=
   S^{01}(p_1, p_2)\,
    \big|Y_{p_2} \chi^{\dot{\alpha}}_{p_1} \big\rangle,\\
    &\mathbf{S}\,\big|\chi^{\dot{\alpha}}_{p_1} \bar{Z}_{p_2} \big\rangle=
    \bar{S}^{01}(p_1, p_2)\,
    \big|\bar{Z}_{p_2}\chi^{\dot{\alpha}}_{p_1} \big\rangle \,.
    \end{aligned}
    \qquad \text{(massive-massless)}
\end{equation}

\begin{equation}
\label{eq:masslessmasslessnorm}
\mathbf{S}\,\big|\chi^{\dot{\alpha}}_{p_1} \chi^{\dot{\beta}}_{p_2} \big\rangle= S^{00}(p_1, p_2) \big|\chi^{\dot{\beta}}_{p_2} \chi^{\dot{\alpha}}_{p_1} \big\rangle .
\qquad \text{(massless-massless)}
\end{equation}
After adding the labels L and R to the Bethe--Yang equations derived in~\cite{Seibold:2022mgg}, according to the particle masses, we obtain the following set of equations.

\paragraph{Left particles.}
\begin{equation}
\label{eq: Bethe left start}
\begin{split}
    1 = e^{i p_i L} &\prod_{\substack{j=1 \\  j \neq i}}^{N_1} S^{11}_{su}(p_i, p_j) \prod_{j=1 }^{N_{\bar 1}} \Tilde S^{11}_{su} (p_i , p_j) \prod_{\dot{\alpha}=1}^{2} \prod_{j=1}^{N^{(\dot{\alpha})}_0}  S^{10}(p_i, p^{(\dot{\alpha})}_j) \prod_{\alpha =1}^2 \prod_{j=1}^{N_y^{(\alpha)}} \overline{S}^{1y}(p_i, y_j^{(\alpha)}) \,.
    \end{split}
\end{equation}
\paragraph{Right particles. }
\begin{equation}
\label{eq: Bethe right start}
\begin{split}
    1 = e^{i p_i L} &\prod_{j = 1}^{N_1} \Tilde S^{11}_{sl}(p_i,p_j) \prod_{\substack{j=1 \\  j \neq i}}^{N_{\bar 1}} S^{11}_{sl} (p_i,p_j) \prod_{\dot{\alpha}=1}^{2} \prod_{j=1}^{N^{(\dot{\alpha})}_0}  \bar{S}^{10}(p_i, p^{(\dot{\alpha})}_j) \prod_{\alpha =1}^2 \prod_{j=1}^{N_y^{(\alpha)}} S^{1y}(p_i, y_j^{(\alpha)}) \,.
    \end{split}
\end{equation} 
\paragraph{Massless particles.}
\begin{equation}
\begin{split}
\label{eq: Bethe massless start}
    1 = e^{i p^{(\dot{\alpha})}_i L} & \prod_{j=1}^{N_1} S^{01}(p^{(\dot{\alpha})}_i, p_j) \prod_{j=1}^{N_{\bar{1}}} \bar{S}^{01}(p^{(\dot{\alpha})}_i, p_j)\\
    &\prod_{\dot{\beta}=1}^2 \prod_{\substack{j=1 \\  (j, \dot{\beta}) \neq (i, \dot{\alpha}) }}^{N^{(\dot{\beta})}_0} S^{00}(p^{(\dot{\alpha})}_i,p^{(\dot{\beta})}_j) \prod_{\alpha=1}^2   \prod_{j = 1}^{N_y^{(\alpha)}}  S^{0y}(p^{(\dot{\alpha})}_i,y_j^{(\alpha)}) \,.
\end{split}
\end{equation} 
\paragraph{Auxiliary roots.}
\begin{equation}
\label{eq: Bethe auxiliary start}
    e^{i\frac{P}{2}} = \prod_{j = 1}^{N_1} \overline{S}^{y1}(y_i^{(\alpha)},p_j) \prod_{j =1}^{N_{\bar 1}} S^{y1}(y_i^{(\alpha)},p_j) \prod_{\dot{\alpha}=1}^2 \prod_{j=1}^{N^{(\dot{\alpha})}_0} S^{y 0} (y_i^{(\alpha)} , p^{(\dot{\alpha})}_j)
\end{equation} 
In the equations above we used the conventions of~\cite{Seibold:2022mgg} indicating with $N_1$ the number of left excitations and with $N_{\bar{1}}$ the number of right excitations. In addition there are $N^{(\dot{1})}_0+N^{(\dot{2})}_0$ massless particles; these particles carry an external $su(2)_{\circ}$ index $\dot{\alpha}$~\cite{Borsato:2014exa}. Finally, there are two types of auxiliary roots transforming in another fundamental representation of a different $su(2)$ algebra, dubbed $su(2)_{\bullet}$ and carrying therefore an index $\alpha=1, \, 2$. To highlight the difference  between the ``massless'' $su(2)_\circ$ and the ``auxiliary'' $su(2)_\bullet$ we use undotted and dotted Greek indices, respectively, to write down their representations.

The S-matrix elements entering the normalization of the highest weight states in~\eqref{eq:massivenorm}, \eqref{eq:massivemasslessnorm} and~\eqref{eq:masslessmasslessnorm} are only partially known in the full theory. Indeed they contain the dressing factors of the model,
whose derivation is still a work in progress; the massive factors have been proposed in~\cite{Frolov:2024pkz,Frolov:2025uwz} (see also~\cite{OhlssonSax:2023qrk}) but the mixed-mass and massless ones are, for now, unknown. However, if we restrict to the relativistic limit, the construction of all dressing factors significantly simplifies and was completed in~\cite{Frolov:2023lwd}, as we will summarise in the next section.  The various auxiliary S-matrices $S^{y1}, \overline{S}^{y1}, S^{y0}$  are independent from the dressing factors and hence completely known. We list them below in terms of the Zhukovsky variables~\eqref{eq:zhukovsky_massive} and~\eqref{eq:zhukovsky_massless}:
\begin{equation}
\label{eq:aux_S_matrices}
    \begin{aligned}
    \overline{S}^{1y}(p, y)&=\frac{1}{\overline{S}^{y 1}(y, p)}= e^{-\frac{i}{2}p} \frac{x^+_{\text{\tiny L}}-y}{x^-_{\text{\tiny L}} - y} \,,\\
    S^{1y}(p, y) &=\frac{1}{S^{y1}(y, p)}= e^{+\frac{i}{2}p} \frac{1-x^-_{\text{\tiny R}}y}{1-x^+_{\text{\tiny R}} y} \,,\\
    S^{0 y} (p , y)&=\frac{1}{S^{y 0} (y , p)}=e^{-\frac{i}{2}p} \frac{x^+ - y}{x^- - y} \,.
    \end{aligned}
\end{equation}
We highlight that differently from~\cite{Seibold:2022mgg}, where the Bethe--Yang equations were derived for even winding sectors only, here we keep track of whether we are in the even or odd winding sector by introducing a factor $e^{i \frac{P}{2}}$ on the LHS of~\eqref{eq: Bethe auxiliary start}. Here $P$ is the total worldsheet momentum and due to the level matching condition must be quantised as
\begin{equation}
P=2 \pi m \quad , \quad m \in \mathbb{N} \,,
\end{equation}
where $m$ is the winding number given by wrapping the string around an angular direction of S$^3$. 
For even $m$ it holds that $e^{i\frac{P}{2}}=1$, while for odd winding number $e^{i\frac{P}{2}}=-1$. Hence, in the string model (as opposed to mirror model) Fermions have periodic boundary conditions in the even winding sector and anti-periodic boundary conditions in the odd winding sector. This changes the periodicity condition in equation~\eqref{eq: Bethe auxiliary start}, as explained in~\cite{Arutyunov:2007tc}.

\subsection{Review of the relativistic limit}

The dispersion relation for the different particles composing the spectrum of the full model is given by~\cite{Lloyd:2014bsa}
\begin{equation}
\label{eq:dispersion}
    E(m,p)= \sqrt{\left(m+\frac{k}{2\pi}p\right)^2+4h^2\sin^2\frac{p}{2}}\,,
\end{equation}
where $m=+1$ and $m=-1$ correspond to massive particles of left and right types, respectively. Representations with $m=0$ are instead associated with `massless' particles. In addition we have infinitely many bound states with $m=\pm2,\, \pm3, \, \dots$ obtained by fusing the fundamental representations.
The relativistic limit considered in~\cite{Frolov:2023lwd} was defined as follows: first, we take $h \ll1$ and we find the minimum of the dispersion relation at the leading order in $h$, which is
\begin{equation}
\label{eq:pminimum}
    p_{\text{min}} = -\frac{2\pi m}{k}\,.
\end{equation}
Then we expand the dispersion relation in~\eqref{eq:dispersion} around this minimum with fluctuations of order $h$.
Introducing the mass
\begin{equation}
\label{eq:mu_definition}
    \mu_m \equiv 2 h \,\left|\sin\frac{\pi m}{k}\right|
\end{equation}
and the following parameterisation for the fluctuations around the minimum
\begin{equation}
\label{eq:pminimum_theta1}
    q_m(\theta) \equiv \mu_m \,\sinh \theta \,,
\end{equation}
then the energies and the momenta of the different particles take the following form
\begin{equation}
\label{eq:rel_lim_p_and_E}
    p_m(\theta) = -\frac{2\pi m}{k}+ \frac{2\pi}{k}q_m(\theta) \,, \qquad  E_m(\theta)= \mu_m \,\cosh\theta \,.
\end{equation}
In the expressions above $\theta$ is identified with a rapidity and the quantities $E_m(\theta)$ and $q_m(\theta)$ are the on-shell energy and momentum of a relativistic particle of mass $\mu_m$:
\begin{equation}
E_m(\theta)^2 - q_m(\theta)^2=\mu^2_m\,.
\end{equation}

By applying the same limit on the Zhukovsky variables in~\eqref{eq:zhukovsky_massive} we obtain the following relations
\begin{equation}
\label{relativistic_limit_XLPM_massive}
    x_{\L}^\pm =
  -  e^{+\theta \mp\frac{i \pi}{k}} \hspace{3mm}, \hspace{3mm} x_{\R}^\pm = e^{-\theta \pm\frac{i \pi }{k}}\,.
\end{equation}
The same limit can be applied to the bound states with arbitrary $m \ne 0 \mod k$, as shown in~\cite{Frolov:2023lwd} and one obtains
\begin{equation}
\label{relativistic_limit_XLPM_gen_m}
x_{\L}^\pm = - \text{sgn} \bigl( \frac{\pi m}{k} \bigl)
   e^{+\theta \mp\frac{i \pi m}{k}} \hspace{3mm}, \hspace{3mm} x_{\R}^\pm = - \text{sgn} \bigl( \frac{\pi m}{k} \bigl)
   e^{-\theta \mp\frac{i \pi m}{k}}\,.
\end{equation}
The fundamental left and right Zhukovsky variables are obtained by substituting respectively $m=+1$ and $m=-1$ in the relations above.

The case $m=0 \ \mod \ k$ is special since for these values of $m$ the dispersion relation is non-analytic at the point $p=-\frac{2 \pi m}{k}$ and we need to distinguish two separate branches of the kinematics, which we label by `$+$' (chiral) and `$-$' (antichiral). These branches correspond to massless particles moving to the right and to the left respectively. For $m=0$ we introduce the following parameterizations for chiral and antichiral massless particles
\begin{equation}
\label{eq:p0_massless_theta}
p^{(\pm)}_0(\theta)= \frac{2 \pi}{k} q^{(\pm)}_0(\theta)\,, \qquad q^{(\pm)}_0(\theta) = \pm h e^{\pm \theta}\,.
\end{equation}
Then the dispersion on the two kinematical branches becomes
\begin{equation}
\label{eq:massless-relativistic}
    E_0^{(\pm)}(\theta)=h\, e^{\pm\theta}\,.
\end{equation}
Applying this limit on the Zhukovsky variables in~\eqref{eq:zhukovsky_massless} we obtain different scaling behaviours depending on the chirality
\begin{equation}
\label{relativistic_limit_XLPM_massless}
x^\pm=
\begin{cases}
\frac{\pi ^2 h^2}{k^2} \left(-\frac{k}{\pi  h}\pm i e^{-\theta}\right) \quad \text{for}\ \ p<0\,\\
\frac{k}{\pi  h}\pm i e^{+\theta} \quad \hspace{15mm} \text{for} \ \ p>0\,.
\end{cases}
\end{equation}
Due to the different behaviours between massive and massless Zhukovsky variables, the massless and massive particles decouple, as well as chiral and antichiral massless particles. This will simplify drastically the Bethe equations and subsequently the TBA equations, as we will see in the next sections.

\subsection{Symmetries in the relativistic limit}
The S~matrix of the string model is fixed, up to the dressing factors, by the algebra appearing in the lightcone gauge-fixed model~\cite{Borsato:2012ss,Lloyd:2014bsa}. This is a subalgebra of $psu(1,1|2)_L\oplus psu(1,1|2)_R$ which gets centrally extended by two charges $\mathbf{C}, \mathbf{C}^\dagger$ (similarly to what happens for $AdS_5\times S^5$ superstrings~\cite{Beisert:2005tm, Arutyunov:2006ak}). In the string mode we have
\begin{equation}
\label{eq:lcsymmetries}
\begin{aligned}
    \{\mathbf{Q}^\alpha,\;\mathbf{S}_\beta\}=&\;\delta^{\alpha}_\beta\,\mathbf{H}\,,&\qquad\{\mathbf{Q}^\alpha,\;\widetilde{\mathbf{Q}}_\beta\}=&\;\delta^{\alpha}_\beta\,\mathbf{C}\,,&\\
    \{\widetilde{\mathbf{Q}}_\alpha,\;\widetilde{\mathbf{S}}^\beta\}=&\;\delta_{\alpha}^\beta\,\widetilde{\mathbf{H}}\,,&\qquad\{\mathbf{S}_\alpha,\;\widetilde{\mathbf{S}}^\beta\}=&\;\delta_{\alpha}^\beta\,\mathbf{C}^\dagger\,,&
\end{aligned}
\end{equation}
where the supercharges without tildes belong to $psu(1,1|2)_L$ and the ones with tildes to $psu(1,1|2)_R$. More specifically, originally the $\mathbf{Q}^\alpha$s and $\widetilde{\mathbf{Q}}_\alpha$s have the interpretation of supersymmetry generators, and the $\mathbf{S}_\alpha$s  and $\widetilde{\mathbf{S}}^\alpha$s of superconformal generators. The labels take values
\begin{equation}
    \alpha=1,2\,,\qquad\beta=1,2\,,
\end{equation}
and in total we have eight real supercharges.
The eigenvalues of the central charges label irreducible representations. For one-particle states we have
\begin{equation}
\begin{aligned}
    \left(\mathbf{H}+\widetilde{\mathbf{H}}\right)|m,p\rangle =&\; E(m,p)\,|m,p\rangle,&\qquad
    \mathbf{C}\;|m,p\rangle =&\; \frac{ih}{2}\left(e^{+ip}-1\right)|m,p\rangle,&\\
    \left(\mathbf{H}-\widetilde{\mathbf{H}}\right)|m,p\rangle =& \left(\frac{k}{2\pi}p+m\right)|m,p\rangle,&\qquad
    \mathbf{C}^\dagger|m,p\rangle =&\; \frac{ih}{2}\left(1-e^{-ip}\right)|m,p\rangle.&
\end{aligned}
\end{equation}
On multiparticle states, there is a non-trivial Hopf algebra structure encoded in the coproduct
\begin{equation}
    \Delta\big(\mathbf{Q}^\alpha\big) = \mathbf{Q}^\alpha\otimes \mathbf{1}+
    e^{\frac{i}{2}\mathbf{p}}\otimes \mathbf{Q}^\alpha\,,\qquad
    \Delta\big(\mathbf{S}_\alpha\big) = \mathbf{S}_\alpha\otimes \mathbf{1}+
    e^{-\frac{i}{2}\mathbf{p}}\otimes \mathbf{S}_\alpha\,,
\end{equation}
where we schematically indicated by $\mathbf{Q}^\alpha$ (respectively, $\mathbf{S}_\alpha$) any of the supersymmetry (respectively, superconformal) generators in either copy of $psu(1,1|2)$ and $\mathbf{p}$ is the worldsheet momentum operator.
These relations are sufficient to fix the scattering matrix up to the dressing factors~\cite{Lloyd:2014bsa}. In fact, as explained in detail in~\cite{Sfondrini:2014via}, it is possible to decompose
\begin{equation}
\begin{aligned}
    \mathbf{Q}^1=\mathbf{Q}\otimes\mathbf{1}\,,\quad
    \mathbf{Q}^2=\mathbf{1}\otimes\mathbf{Q}\,,\qquad
    \mathbf{S}_1=\mathbf{S}\otimes\mathbf{1}\,,\quad
    \mathbf{S}_2=\mathbf{1}\otimes\mathbf{S}\,,\\
    \widetilde{\mathbf{Q}}_1=\widetilde{\mathbf{Q}}\otimes\mathbf{1}\,,\quad
    \widetilde{\mathbf{Q}}_2=\mathbf{1}\otimes\widetilde{\mathbf{Q}}\,,\qquad
   \widetilde{\mathbf{S}}_1=\widetilde{\mathbf{S}}\otimes\mathbf{1}\,,\quad
    \widetilde{\mathbf{S}}_2=\mathbf{1}\otimes\widetilde{\mathbf{S}}\,,
\end{aligned}
\end{equation}
and similarly decompose the representation under which particles transform. Then, we may use the ``auxiliary'' algebra generated by $\mathbf{Q},\mathbf{S},\widetilde{\mathbf{Q}},\widetilde{\mathbf{S}}$ to determine an ``auxiliary'' S~matrix, while the complete ~matrix is given by the tensor product of two such auxiliary blocks.

Let us now see what happens in the relativistic limit. We find the eigenvalues
\begin{equation}
\begin{aligned}
    \mathbf{H}\;|m,p\rangle\to&\; \frac{\mu}{2}e^{+\theta}|\mu,\theta\rangle\,,&
    \qquad
    \mathbf{C}\;|m,p\rangle\to&\; \frac{\mu}{2}e^{-\frac{i\pi m}{k}}|\mu,\theta\rangle\,,&\\
    \widetilde{\mathbf{H}}\,|m,p\rangle\to&\; \frac{\mu}{2}e^{-\theta}|\mu,\theta\rangle\,,&
    \qquad
    \mathbf{C}^\dagger|m,p\rangle\to&\; \frac{\mu}{2}e^{+\frac{i\pi m}{k}}|\mu,\theta\rangle\,,&
\end{aligned}
\end{equation}
where we assumed for simplicity $0< m <k$. The coproduct also remains nontrivial, albeit it is no longer momentum dependent:
\begin{equation}
\label{eq:twistedcoproduct}
    \Delta\big(\mathbf{Q}^\alpha\big) = \mathbf{Q}^\alpha\otimes \mathbf{1}+
    e^{-\frac{i\pi m}{k}}\;\mathbf{1}\otimes \mathbf{Q}^\alpha\,,\qquad
    \Delta\big(\mathbf{S}_\alpha\big) = \mathbf{S}_\alpha\otimes \mathbf{1}+
    e^{+\frac{i\pi m}{k}}\;\mathbf{1}\otimes \mathbf{S}_\alpha\,.
\end{equation}
A similar limit can also be taken for massless representations. In that case, only the chiral or antichiral Hamiltonian ($\mathbf{H}$ or $\widetilde{\mathbf{H}}$, respectively) are non-zero; note that the central charges $\mathbf{C},\mathbf{C}^\dagger$ also vanish on this representation, and as a result the coproduct~$\Delta$ is trivial.
The form of the central charges is precisely that of a relativistic model with $\mathcal{N}=(2,2)$ supersymmetry. However, here we have \textit{too many supercharges}: indeed, if restricted to one copy of the supercharges $\mathbf{Q},\mathbf{S},\widetilde{\mathbf{Q}},\widetilde{\mathbf{S}}$ --- equivalently, restricting to $\alpha=\beta=1$ in~\eqref{eq:lcsymmetries} --- we would precisely reproduce the $\mathcal{N}=(2,2)$ relations, including the twisted coproduct~\eqref{eq:twistedcoproduct} which plays a crucial role in the S-matrix construction of~\cite{Fendley:1991ve,Fendley:1992dm}. Instead, we have \textit{two distinct $\mathcal{N}=(2,2)$ algebras}, labeled by $\alpha=1,2$ which \textit{have the same central elements}. Remark that this is not the $\mathcal{N}=(4,4)$ supersymmetry algebra, which would have a non-abelian R-symmetry. Rather, here we are ``squaring'' the $\mathcal{N}=(2,2)$ symmetry.
In any case, the fact that we match the symmetries and coproduct of~\cite{Fendley:1991ve,Fendley:1992dm} has important consequences: since the whole S-matrix can be found from tensoring two representations of the smaller auxiliary algebra of the $\mathbf{Q},$ $\mathbf{S},$ $\widetilde{\mathbf{Q}},$ $\widetilde{\mathbf{S}}$, it follows that the full S~matrix of the FPS model is obtained by tensoring two copies of the S-matrix found by FI~\cite{Fendley:1991ve,Fendley:1992dm} (in the massive sector). We hence expect the TBA equations will also be closely related to those of~\cite{Fendley:1991ve,Fendley:1992dm}, though not identical as we shall see --- nor will the UV central charge be twice the one of~\cite{Fendley:1991ve,Fendley:1992dm}.

\subsection{Limit of the Bethe--Yang equations}
\label{sec:Bethe--Yang equations after the limit}

Now we evaluate the relativistic limit of the Bethe--Yang equations. 
First, we define
\begin{equation}
N^{(\dot{\alpha})}_0= N^{(+, \dot{\alpha})}_0 + N^{(-, \dot{\alpha})}_0
\end{equation}
to be the total number of massless particles for a given index $\dot{\alpha}=1, 2$. Among them there are $N^{(+, \dot{\alpha})}_0$ massless particles moving to the right and $N^{(-, \dot{\alpha})}_0$ massless particles moving to the left. In~\cite{Frolov:2023lwd} it was observed that massive and massless particles decouple, and so do chiral and antichiral massless particles. Correspondingly, the Bethe equations will split into three sectors coupled only by the phase shifts due to constant S-matrix elements. These sectors are composed of massive particles, massless chiral particles and massless antichiral particles.
Moreover, the total contribution due to the constant shifts can be absorbed in an overall twist of the Bethe--Yang equations, dependent on the charges of the state (or equivalently, on the number of particles that appear in it) but not on the momenta.
Below we derive these equations starting from those  written in section~\ref{sec:Bethe_Eq_before_limit}.

\paragraph{Relativistic limit of the auxiliary roots.}

We split the auxiliary roots into three groups depending on their behaviour in the relativistic limit (in particular, as $h\to0$). We call these groups `massive', `chiral' and `antichiral' because as we will see each of them will only couple to one type of momentum carrying roots. Moreover, as we will see auxiliary roots with any different behaviour would completely drop from the Bethe--Yang equations. This strongly suggests that it should be possible to split all auxiliary roots (for a fixed index $\alpha=1, 2$) as
\begin{equation}
N^{(\alpha)}_y = N^{(\alpha)} + N^{(+,\alpha)} + N^{(-,\alpha)} \,,
\end{equation}
where $N^{(\alpha)}$,  $N^{(+,\alpha)}$ and $N^{(-,\alpha)}$ refer to the numbers of massive, chiral and antichiral auxiliary roots respectively. Indeed, if we allowed for some of the auxiliary roots to completely decouple, we would obtain additional degeneracies in the spectrum which we find hard to justify physically.
Hence, we assume that we have the following three possibilities:
\begin{enumerate}
    \item A massive auxiliary root $y_j^{(\alpha)}$ such that, as $h\to0$,
    \begin{equation}
\label{eq:massive_aux_root_par}
y_j^{(\alpha)} \equiv e^{\theta^{(\alpha)}_j} \, ,\hspace{3mm} j=1, \dots, N^{(\alpha)} \,.
\end{equation}
\item A chiral auxiliary root $y_j^{(\alpha)}$ such that, as $h\to0$,
    \begin{equation}
\label{eq:chiral_aux_root_par}
y_j^{(\alpha)} \equiv \frac{k}{\pi h} - e^{\theta^{(\alpha)}_j} \, ,\hspace{3mm} j=1, \dots, N^{(+,\alpha)} \,.
\end{equation}
\item An antichiral auxiliary root $y_j^{(\alpha)}$ such that, as $h\to0$,
    \begin{equation}
\label{eq:antichiral_aux_root_par}
y_j^{(\alpha)} \equiv -\frac{\pi h}{k} + \frac{\pi^2 h^2}{k^2} e^{-\theta^{(\alpha)}_j} \, ,\hspace{3mm} j=1, \dots, N^{(-,\alpha)} \,.
\end{equation}
\end{enumerate}
Each of these behaviours is related to the scaling of the momentum carrying roots in the relativistic limit, and will result non-trivial auxiliary S matrices in the limit.
Let us consider the S-matrix $S^{y 0} (y, p)$ describing the interaction between an auxiliary root $y$ with a chiral massless particle. At the leading order in $h$ the momentum of the massless particle is null and plugging~\eqref{relativistic_limit_XLPM_massless} into the last row of~\eqref{eq:aux_S_matrices} we obtain
\begin{equation}
S^{y 0} (y, p) = e^{i\frac{p}{2}} \frac{x^- - y}{x^+ - y} \to \frac{\frac{k}{\pi h}-ie^\theta- y}{\frac{k}{\pi h}+ie^\theta-y} \,.
\end{equation}
If $y$ was parameterised as in~\eqref{eq:massive_aux_root_par} or~\eqref{eq:antichiral_aux_root_par} we immediately see that the S-matrix becomes equal to one in the limit $h\to 0$. However, if $y$ was parameterised as in~\eqref{eq:chiral_aux_root_par} then we obtain
\begin{equation}
S^{y 0} (y_j^{(\alpha)}, p_i) \to \tanh \bigl( \frac{\theta_j^{(\alpha)}-\theta_i}{2} - \frac{i \pi}{4} \bigl)  \,.
\end{equation}
In a similar manner if we parameterise the antichiral auxiliary roots as in~\eqref{eq:antichiral_aux_root_par} then the scattering of an antichiral auxiliary root with an antichiral massless particle is
\begin{equation}
S^{y 0} (y_j^{(\alpha)}, p_i) \to \tanh \bigl( \frac{\theta_i-\theta_j^{(\alpha)}}{2} + \frac{i \pi}{4} \bigl)  \,.
\end{equation}

The scattering of massive particles with auxiliary roots of type~\eqref{eq:massive_aux_root_par} is also easily obtained by substituting the expressions~\eqref{relativistic_limit_XLPM_massive} into the first two lines of~\eqref{eq:aux_S_matrices}.
Defining the rapidity difference $\theta \equiv \theta_i - \theta_j$ (in the following we also label this difference by $\theta_{ij}$), they are given by
\begin{equation}
\label{eq:Ups_1_km1}
\begin{aligned}
    &\overline{S}^{1y}(p_i, y_j^{(\alpha)}) \to \Upsilon(1; \theta)\,,\\
    &S^{1y}(p_i, y_j^{(\alpha)}) \to - \Upsilon(k-1; \theta) \,,
\end{aligned}
\end{equation}
where we define 
\begin{equation}
\label{eq:aux_phys_dressing_m}
\Upsilon(m; \theta) \equiv \frac{\cosh \bigl(\frac{\theta}{2}-m\frac{i \pi}{2k} \bigl)}{\cosh \bigl(\frac{\theta}{2}+m\frac{i \pi}{2k} \bigl)} \,.
\end{equation}
In a similar way, it is easy to check that any other scattering of auxiliary and main roots (massive with massless, chiral with anti-chiral) results in a trivial auxiliary S~matrix in the limit.
This fact indicates that in the relativistic limit the Bethe--Yang equations for massive-, chiral- and antichiral-particles decouple, as expected from the results of~\cite{Frolov:2023lwd}.

\paragraph{Relativistic limit of the Bethe equations for physical particles.}

Let us now review the the dressing factors of the relativistic model  which have been found in~\cite{Frolov:2023lwd}. Following that paper, we split it into two parts: a minimal solution of the crossing equations, and a ``CDD''~\cite{Castillejo:1955ed} factor which accounts for the bound-state structure of the model and solves the homogeneous crossing equation.
The minimal solutions for massive particles are
\begin{equation}
\label{minimal_sigma_ratio_of_R_functions}
    \sigma(m_1,m_2;\theta)^{-2} =  \frac{R\left(\theta-\frac{i\pi(m_1+m_2)}{k}\right)^2\,R\left(\theta+\frac{i\pi(m_1+m_2)}{k}\right)^2}{R\left(\theta-\frac{i\pi(m_1-m_2)}{k}\right)^2 \,R\left(\theta+\frac{i\pi(m_1-m_2)}{k}\right)^2} \,,
\end{equation}
where $R(\theta)$ can be expressed in terms of $\Gamma$-functions or Barnes~$G$-function%
\footnote{The $G$-function obeys $G(z+1)=\Gamma(z)\,G(z)$ with $G(1)=1$. The function $\psi(z)$ is the Digamma function, defined by $\psi(z)=\tfrac{\de}{\de z}\log\Gamma(z)$.}
\begin{equation}
\label{R_function_definition}
     R(\theta)\equiv \frac{G(1- \frac{\theta}{2\pi i})}{G(1+ \frac{\theta}{2\pi i}) } =   \left(\frac{e}{2\pi}\right)^{+\frac{\theta}{2\pi i}}\prod_{\ell=1}^\infty \frac{\Gamma(\ell+\frac{\theta}{2\pi i})}{\Gamma(\ell-\frac{\theta}{2\pi i})}\,e^{-\frac{\theta}{\pi i}\,\psi(\ell) }
\end{equation}
and $m_1, m_2 \in \{1, \dots, k-1\}$. The value $m=1$ is associated with left particles while the value $m=k-1$ is associated with right particles. The intermediate quantum numbers $m=2, \dots, k-2$ are associated with bound states, which can be created by fusing left or right particles. We will show how the fusion procedure works at the level of the Bethe--Yang equations in one moment.

The dressing factors in~\eqref{minimal_sigma_ratio_of_R_functions} need to be further dressed with CDD-like factors of the form\footnote{Such $\Phi(m_1,m_2;\theta)$  satisfies the homogeneous crossing equation $\Phi(m_1,m_2;\theta)=\Phi(k-m_1,m_2;i \pi-\theta)$.} 
\begin{equation}
    \label{general_formula_for_CDD_factor}
     \Phi(m_1,m_2;\theta)= [m_1+m_2]_\theta [m_1+m_2-2]^2_\theta 
     \dots [|m_1-m_2|+2]^2_\theta [|m_1-m_2|]_\theta \,,
\end{equation}
where
\begin{equation}
\label{definition_building_block}
    [m]_\theta\equiv \frac{\sinh{\Bigl( \frac{\theta}{2} + \frac{i \pi m}{2k} \Bigl)}}{\sinh{\Bigl( \frac{\theta}{2} - \frac{i \pi m}{2k} \Bigl)}} \,.
\end{equation}
These additional factors take into account the poles expected by the propagation of bound states and correspond to minimal S-matrices of the $A^{(1)}_{k-1}$ series of affine Toda theories (see e.g.~\cite{Braden:1989bu}). 

The solutions for the dressing factors associated with the scattering of massless particles in the relativistic limit were also found in~\cite{Frolov:2023lwd}. We can use them to express the various S-matrix elements coupling the momentum-carrying modes in the Bethe--Yang equations. 
We have the following S-matrix elements:
\begin{enumerate}
    \item left-left and right-right:
\begin{equation}
\begin{split}
&S^{11}_{su}(p_1, p_2) \to \Phi(1,1;\theta_{12})\sigma(1,1;\theta_{12})^{-2} \,, \\
&S^{11}_{sl}(p_1, p_2) \to \Phi(k-1,k-1;\theta_{12})\sigma(k-1,k-1;\theta_{12})^{-2} \,;
\end{split}
\end{equation}
\item left-right and right-left:
\begin{equation}
\begin{split}
&\tilde{S}^{11}_{su}(p_1, p_2) \to \Phi(1,k-1;\theta_{12})\sigma(1,k-1;\theta_{12})^{-2} \,, \\
&\tilde{S}^{11}_{sl}(p_1, p_2) \to \Phi(k-1,1;\theta_{12})\sigma(k-1,1;\theta_{12})^{-2} \,;
\end{split}
\end{equation}
    \item chiral-antichiral and antichiral-chiral massless particles:
    \begin{equation}
S^{00}(p_1,p_2) \to 1 \,;
    \end{equation}
    \item chiral-chiral and antichiral-antichiral massless particles:
    \begin{equation}
    \label{eq:massless_S_mat}
S^{00}(p_1,p_2) \to \Sigma^{00}(\theta_{12})=a(\theta_{12}) \Bigl(\frac{R(\theta_{12} - i \pi) R(\theta_{12} + i \pi)}{R^2(\theta_{12})} \Bigl)^2 \,,
    \end{equation}
where
\begin{equation}
a(\theta)= -i \tanh \Bigl(\frac{\theta}{2} - i \frac{\pi}{4} \Bigl) \,;
\end{equation}
    \item chiral-left and chiral-right: 
    \begin{equation}
 S^{01}(p_1, p_2) \to e^{-\frac{i \pi}{k}} \hspace{3mm}, \hspace{3mm} \bar{S}^{01}(p_1, p_2) \to e^{-\frac{i \pi}{k} (k-1)} \,\,;
    \end{equation}
    \item left-antichiral and right-antichiral:
    \begin{equation}
 S^{10}(p_1, p_2) \to 1 \hspace{3mm}, \hspace{3mm} \bar{S}^{10}(p_1, p_2) \to 1 \,\,.
    \end{equation}
\end{enumerate}

\subsection{String hypothesis for the FPS model}

Taking the limit of Bethe--Yang equation is a convenient way to avoid bothering with the diagonalisation of the relativistic S~matrix (which we anyway sketch in appendix~\ref{app:first_appendix}, the result is identical). However, the limit is not truly well-defined at the level of the Bethe--Yang equations. To see this, note that the auxiliary S~matrices as well as those of the momentum carrying modes admit finite limit as $h\to0$. Denoting such S-matrices collectively as $S(p_i,p_j)$, we therefore find that in the limit the Bethe--Yang equations for a left particle give
\begin{equation}
    1=e^{i p_i L}\prod_j S(p_i,p_j)\qquad
    \longrightarrow\qquad
    1=e^{-\frac{2\pi i L}{k}}e^{\frac{2\pi i L}{k}q_1(\theta_i)}\prod_j S(\theta_{ij})\,.
\end{equation}
Because the momentum is $q_1(\theta)=\mu_1 \sinh\theta$ with $\mu_1=\mathcal{O}(h)$, to obtain a sensible set of Bethe--Yang equations we have to assume that $L=\mathcal{O}(h^{-1})$ as $h\to0$. However, this makes the factor $e^{-\frac{2\pi i L}{k}}$ oscillate rapidly, and a very ad-hoc procedure for the limit would be needed (for instance, ensuring that $L=\mathcal{O}(h^{-1})$ and that is a multiple of~$k$). This is not all: physical states of the original non-relativistic worldsheet model had to satisfy the level-matching constraint. After the limit, this constraint reads
\begin{equation}
P = \frac{2 \pi}{k} \bigl( N_{\bar{1}} - N_1 \bigl)+\sum_j q_j(\theta_j) =0~\text{mod}~2\pi\,.
\end{equation}
Again, bearing in mind that $q_j(\theta_j)=\mathcal{O}(h)$ this imposes two conditions: first that for a physical state $N_{\bar{1}}=N_1$ mod~$k$, second that the total momentum vanishes.
However, in the original string theory the level-matching constraint would play no role in the derivation of the \textit{mirror} TBA (because there is no such constraint \textit{in the mirror model}); it would only matter in the analysis of the excited states.

All this shows that we cannot really use the Bethe--Yang equations of the string model to obtain ``the relativisitic limit of the spectrum''. Instead, the more natural logic is that the relativistic limit may be taken \textit{at the level of the S~matrix} (or even more fundamentally, of the symmetry algebra), which gives the FPS model. Starting from such a relativistic model one can consider the (thermal or mirror) TBA equations of the model. In practice, the thermal and mirror equations would only differ by a chemical potential in the free energy --- related to the boundary conditions of the fermions.  

Therefore, we will henceforth treat the mass $\mu$ and hence~$q(\theta)$ as quantities of order one, without any reference to~$h$ (which has indeed dropped out of our equations). We let $R$ be the spatial volume for the relativistic model, so that the Bethe equations take the schematic form
\begin{equation}
    1=e^{i q(\theta_i)\,R}\prod_j S(\theta_{ij})\,.
\end{equation}

It is also worth commenting on the various constant twists that emerge from e.g.\ the mixed-mass sector of the scattering. Consider for instance the Bethe--Yang equation for a particle with $m=1$. If we kept track of all the various twists we would find 
\begin{multline}
    1 = e^{i q_1(\theta_i)\,R}  \prod^2_{\alpha=1}e^{i\frac{\pi}{k}(N^{(+, \alpha)}-N^{(-, \alpha)})}  \prod^2_{\dot{\alpha}=1}e^{i\frac{\pi}{k} N_0^{(+, \dot{\alpha})}}
    \prod_{\substack{j=1 \\  j \neq i}}^{N_1} \Phi(1,1;\theta_{ij})\sigma(1,1;\theta_{ij})^{-2}\\
    \times\prod_{j=1 }^{N_{\bar 1}} \Phi(1,k-1;\theta_{ij}) \sigma(1,k-1;\theta_{ij})^{-2} \prod_{\alpha =1}^2\prod_{j=1}^{N^{(\alpha)}} \Upsilon(1; \theta_{i}-\theta^{(\alpha)}_{j}) \,.
\end{multline}
The constant phases can be reabsorbed in an overall twist of the boundary conditions depending on the total number of excitations. In the original string theory, these twists have an interpretation in terms of ``frame factors'' of the model, necessary to recover the $psu(1,1|2)_L\oplus psu(1,1|2)_R$ invariance of the spectrum. Since this invariance is destroyed by the relativistic limit%
\footnote{%
To see this, note that to create a symmetry descendant one would act with a $p=0$ of the charges. Such modes are inaccessible from the region $p\approx -2\pi m/k$ where we are. Moreover, creating a descendant would change the energy $E(m,p)$ by one unit --- a quantity parametrically larger than our $\mathcal{O}(h)$ energies.
}
there is no reason to keep track of such factors, that are anyway irrelevant for the purpose of deriving the TBA equations. We therefore simply write equations such as 
\begin{multline}
\label{eq:left_particle_Bethe_Yang_eq}
    1 \sim e^{i q_1(\theta_i)\,R}  
    \prod_{\substack{j=1 \\  j \neq i}}^{N_1} \Phi(1,1;\theta_{ij})\sigma(1,1;\theta_{ij})^{-2}
    \prod_{j=1 }^{N_{\bar 1}} \Phi(1,k-1;\theta_{ij}) \sigma(1,k-1;\theta_{ij})^{-2}\\ 
    \times\prod_{\alpha =1}^2\prod_{j=1}^{N^{(\alpha)}} \Upsilon(1; \theta_{i}-\theta^{(\alpha)}_{j}) \,.
\end{multline}

It is known~\cite{Frolov:2023lwd} that, besides fundamental particles of $m=1$ and $m=k-1$, this model has bound states, so that the full particle spectrum spans $m=1,2,\dots, k-1$, with particles with mass $m$ and $k-m$ being conjugate to each other. The same can be seen from the Bethe--Yang equation, see appendix~\ref{app:BS}, where the Bethe--Yang equations of the bound states are derived by fusion. To summarise that construction, we have 
\begin{enumerate}
    \item a number $N_Q$ of bound states with real momenta obtained by fusing $Q$ particles with $m=1$. 
    \item two sets of $N^{(+, \dot{\alpha})}_0$ chiral massless particles and two sets of $N^{(-, \dot{\alpha})}_0$ antichiral massless particles with flavour $\dot{\alpha}=1, 2$;
    \item two sets of $N^{(\alpha)}$ massive `auxiliary roots' $y^{(\alpha)}$ with flavour $\alpha = 1, 2$;
    \item  two sets of $N^{(+,\alpha)}$ chiral auxiliary roots and two sets $N^{(-,\alpha)}$ of antichiral  auxiliary roots ($\alpha=1, 2$).
\end{enumerate}
From now on we will say that a particle has length $Q$ if can be obtained by fusing a number $Q$ of left fundamental particles.
The complete set of Bethe--Yang equations can be compactly written as follows.

\paragraph{Equations for massive particles.}

\begin{equation}
\label{eq1:BAEq_fund_bound_states}
\begin{split}
1 \sim e^{iR q_{P}(\theta_i)}  \prod_{Q=1}^{k-1}\prod_{j =1}^{N_Q} \Phi(P,Q;\theta_{ij})\sigma(P,Q;\theta_{ij})^{-2} \prod_{\alpha =1}^2 \prod_{j=1}^{N^{(\alpha)}} \Upsilon(P; \theta_i- \theta^{(\alpha)}_j) \,,
\end{split}
\end{equation}
\begin{equation}
\label{eq2:BAEq_fund_bound_states}
1   \sim \prod^{k-1}_{Q=1}\prod_{j = 1}^{N_Q} \Upsilon(Q; \theta^{(\alpha)}_i- \theta_j)  \,.
\end{equation}
In the expression above we labelled by $q_P(\theta_i)$ the momentum of the $i^{\text{th}}$-particle of length $P$, where $i=1, \dots, N_P$.

\paragraph{Equations for chiral particles}
\begin{equation}
\label{eq:Bethe_chiral_rl_bs}
    1 \sim e^{iR q^{(+)}_0(\theta^{\dot{\alpha}}_i)}  \prod_{\dot{\beta}=1}^2   \prod_{j = 1}^{N^{(+, \dot{\beta})}_0}  \Sigma^{00}(\theta^{(\dot{\alpha})}_i - \theta^{(\dot{\beta})}_j) \prod_{\alpha=1}^2   \prod_{j = 1}^{N_y^{(+, \alpha)}}  \coth \Bigl( \frac{\theta^{(\alpha)}_j - \theta^{(\dot{\alpha})}_i}{2} - \frac{i \pi}{4} \Bigl) \,,
\end{equation} 
\begin{equation}
\label{eq:Bethe_auxiliary_chiral_rl_bs}
    1 \sim \prod_{\dot{\alpha}=1}^2 \prod_{j=1}^{N^{(+,\dot{\alpha})}_0} \tanh \bigl( \frac{\theta_i^{(\alpha)}-\theta^{(\dot{\alpha})}_j}{2} - \frac{i \pi}{4} \bigl) \,.
\end{equation}

\paragraph{Equations for antichiral particles}
\begin{equation}
\begin{split}
\label{eq:Bethe_antichiral_rl_bs}
    1 \sim e^{iR q^{(-)}_0(\theta^{\dot{\alpha}}_i)} &\prod_{\dot{\beta}=1}^2   \prod_{j = 1}^{N^{(-, \dot{\beta})}_0}  \Sigma^{00}(\theta^{(\dot{\alpha})}_i - \theta^{(\dot{\beta})}_j) \prod_{\alpha=1}^2   \prod_{j = 1}^{N_y^{(-, \alpha)}}  \coth \Bigl( \frac{\theta^{(\dot{\alpha})}_i-\theta^{(\alpha)}_j}{2} + \frac{i \pi}{4} \Bigl) \,,
\end{split}
\end{equation} 
\begin{equation}
\label{eq:Bethe_auxiliary_antichiral_rl_bs}
    1 \sim \prod_{\dot{\alpha}=1}^2 \prod_{j=1}^{N^{(-,\dot{\alpha})}_0} \tanh \bigl( \frac{\theta^{(\dot{\alpha})}_j - \theta_i^{(\alpha)}}{2} + \frac{i \pi}{4} \bigl) \,.
\end{equation} 

\paragraph{Energy}

The total energy and momentum is
\begin{equation}
\mathcal{E}=\sum^{k-1}_{Q=1} \sum^{N_Q}_{j=1} E_Q (\theta_j) +\sum_{\dot\alpha=1}^2\sum_{j=1}^{N_0^{(+,\dot\alpha)}}E^{(+)}_0(\theta_j)+\sum_{\dot\alpha=1}^2\sum_{j=1}^{N_0^{(-,\dot\alpha)}}E^{(-)}_0(\theta_j)\,.
\end{equation}
Of course, since the Bethe--Yang equations have decoupled, we can also separately consider the energy of massive, chiral, and antichiral particles --- corresponding to the three terms in the above formula.

\paragraph{String hypothesis.}
The excitations listed above are all those which can appear in large-$R$ limit. The construction of the bound states already takes care of the only ``Bethe strings'' of the model, as it is not possible to construct strings involving momentum carrying and auxiliary particles --- much like in the mirror TBA of~\cite{Frolov:2021bwp}. For momentum carrying particles, the rapidity has to necessarily be real to ensure reality of the energy (as well as to solve the equations). For auxiliary roots, we have more freedom. It is not difficult to check that the function $\Upsilon(m; \theta)$ in~\eqref{eq:aux_phys_dressing_m} satisfies the following property
\begin{equation}
\begin{split}
&|\Upsilon(m; \theta)| > 1 \ \text{if} \hspace{9mm} 0<\text{Im}(\theta)< \pi \mod 2 \pi \,,\\
&|\Upsilon(m; \theta)| < 1 \ \text{if} \hspace{4mm} -\pi<\text{Im}(\theta)< 0 \mod 2 \pi \,,\\
&|\Upsilon(m; \theta)| = 1 \ \text{if} \hspace{16mm} \text{Im}(\theta)= 0 \mod \pi \,.
\end{split}
\end{equation}
Hence, by the usual reasoning, in the thermodynamic limit only the last case can contribute. Bearing in mind that momentum-carrying particles have real rapidity, it follows that for auxiliary ones $\text{Im}(\theta^{(\alpha)})=0$ mod~$i \pi$.%
\footnote{Thinking back at the original parametrisation~\eqref{eq:massive_aux_root_par} this is just the statement that the root~$y$ may be positive or negative.}
The same argument can be applied to $\tanh \bigl( \frac{\theta}{2} \pm \frac{i \pi}{4} \bigl)$ to show that auxiliary roots need to be real $\mod i \pi$ also for in the chiral and antichiral equations.

\section{Thermodynamic Bethe Ansatz equations}
\label{sec:TBA}

In this section we derive the TBA equations following a standard route: first, we write the logarithm of the  Bethe--Yang equations and take their thermodynamic limit (introducing integrals and convolutions); then we require that the free energy is stationary at the thermodynamic equilibrium obtaining in this manner the TBA equations.

\subsection{TBA kernels}
\label{sec:equations_for_densities}
To begin with let us introduce the kernels associated with the different scattering matrices of the system. They are the logarithmic derivatives of the various S matrices appearing in the Bethe--Yang equations.
The kernels are defined so that they may be integrated on the real-$\theta$ line in the TBA equations. Recall that in our string hypothesis auxiliary roots allowed both real-rapidity solutions, and solutions with $\text{Im}(\theta)=i\pi$. To accommodate the shift of $i\pi$ we will introduce a distinct kernel for the latter case.
Below we list all the kernels which will be used in the remaining part of this section.

\paragraph{Universal (Cauchy) kernel.}
A fundamental ingredient of the TBA equations is the Cauchy kernel. Here we find convenient to introduce a family of kernels, labeled by~$k$, given by
\begin{equation}
\label{eq:Cauchy_kernel_definition}
s_k(\theta)=\frac{k}{4 \pi}\frac{1}{\cosh \frac{k \theta}{2}}\,.
\end{equation}
The standard Cauchy kernel is the one with $k=2$. For future convenience, we also define the right-inverse $s_k^{-1}$ by
\begin{equation}
(f*s_k^{-1})(\theta)=\lim_{\epsilon\to0^+}\left(f(\theta+\frac{i \pi}{k}- i \epsilon)+f(\theta-\frac{i \pi}{k}+ i \epsilon) \right),
\end{equation}
so that
\begin{equation}
(s_k*s_k^{-1})(\theta)= \delta(\theta)\,.
\end{equation}
The right-inverse is useful for deriving the Y-system equations, as we shall see.

\paragraph{Kernels for massive particles.}
The kernel associated with the scattering of physical massive particles can be written in the following form
\begin{equation}
K^{PQ}(\theta)= \frac{1}{2 \pi i} \frac{d}{d \theta} \log \bigl( \Phi(P,Q;\theta)\sigma(P,Q;\theta)^{-2} \bigl)= K_{\sigma}^{PQ}(\theta)+K_{\Phi}^{PQ}(\theta) \,,
\end{equation}
where
\begin{equation}
\begin{split}
&K_{\sigma}^{PQ}(\theta)=  \frac{1}{2 \pi i} \frac{d}{d \theta} \log \sigma(P,Q;\theta)^{-2}\\
&=- \frac{1}{4 \pi^2} \Bigl[ \bigl( \theta - \tfrac{i \pi(P+Q)}{k}  \bigl) \coth \bigl( \tfrac{\theta}{2} - \tfrac{i \pi(P+Q)}{2k}  \bigl)
+\bigl( \theta + \tfrac{i \pi (P+Q)}{k} \bigl) \coth \bigl( \tfrac{\theta}{2} + \tfrac{i \pi(P+Q)}{2k}  \bigl)\\
&\qquad\qquad-\bigl( \theta - \tfrac{i \pi(P-Q)}{k}  \bigl) \coth \bigl( \tfrac{\theta}{2} - \tfrac{i \pi(P-Q)}{2k}  \bigl)
-\bigl( \theta + \tfrac{i \pi(P-Q)}{k}  \bigl) \coth \bigl( \tfrac{\theta}{2} + \tfrac{i \pi(P-Q)}{2k}  \bigl) \Bigl]
\end{split}
\end{equation}
and
\begin{equation}
\begin{split}
K_{\Phi}^{PQ}(\theta)&=  \frac{1}{2 \pi i} \frac{d}{d \theta} \log \Phi(P,Q;\theta)\\
&= -\langle P+Q \rangle_{\theta} -2 \langle P+Q-2\rangle_{\theta}+\dots-2 \langle |P-Q|-2\rangle_{\theta}-\langle |P-Q| \rangle_{\theta} \,.
\end{split}
\end{equation}
The expression above is written in terms of the building blocks
\begin{equation}
\label{eq:angle_brackets_B_Blocks}
\langle x \rangle_{\theta} \equiv \frac{1}{2 \pi} \frac{\sin \frac{\pi x}{k} }{\cosh{\theta} - \cos \frac{\pi x}{k}  } \,.
\end{equation}
The kernels associated with the scattering between massive particles and auxiliary roots of massive type  are instead given by the following relations 
\begin{equation}
K^{Py}_{\downarrow}(\theta)=K^{yP}_{\downarrow}(\theta) = -\frac{1}{2 \pi i} \frac{d}{d \theta} \log \bigl( \Upsilon(P;\theta) \bigl) = \langle k-P \rangle_{\theta} \,,
\end{equation}
\begin{equation}
K_{\uparrow}^{Py}(\theta)=K_{\uparrow}^{yP}(\theta) = \frac{1}{2 \pi i} \frac{d}{d \theta} \log \bigl( \Upsilon(P;\theta-i \pi) \bigl) = \langle P \rangle_{\theta} \,.
\end{equation}
Following the previous discussion two different kernels for the auxiliary massive particles have been introduced because the auxiliary rapidities solving~\eqref{eq2:BAEq_fund_bound_states} are of the form $\theta^{(\alpha)}_i=\lambda_i^{(\alpha)}$ and $\theta^{(\alpha)}_i=\tilde{\lambda}_i^{(\alpha)}+i \pi$ with $\lambda_i^{(\alpha)}$ and $\tilde{\lambda}_i^{(\alpha)}$ reals.

\paragraph{Kernels for massless particles.}
The S matrix describing the scattering of chiral-chiral physical particles is the same as the S matrix describing the scattering of antichiral-antichiral physical particles. The common kernel is
\begin{equation}
K^{00}(\theta)= \frac{1}{2 \pi i} \frac{d}{d \theta} \log \Sigma^{00}(\theta)= s_2(\theta) + 2 \phi_0 (\theta)\,,
\end{equation}
where $s_2$ is given by~\eqref{eq:Cauchy_kernel_definition} after substituting $k=2$ and $\phi_0$ is the sine-Gordon kernel
\begin{equation}
\label{eq:sine_G_kernel}
\phi_0(\theta)=\frac{1}{2\pi^2} \frac{\theta}{\sinh{\theta}} \,.
\end{equation}
As before we split the chiral auxiliary roots into two sets, having rapidities with imaginary part equal to $0$ and $i \pi$ respectively. The kernels associated with the scattering between chiral massless particles and chiral auxiliary roots are then 
\begin{equation}
K_{\downarrow}^{0+}(\theta)=K_{\downarrow}^{+0}(\theta)= \frac{1}{2 \pi i} \frac{d}{d \theta} \log \tanh \Bigl( \frac{\theta}{2} - \frac{i \pi}{4} \Bigl) = s_2(\theta) \,,
\end{equation}
\begin{equation}
K_{\uparrow}^{0+}(\theta)= K_{\uparrow}^{+0}(\theta)= - \frac{1}{2 \pi i} \frac{d}{d \theta} \log \coth \Bigl( \frac{\theta}{2} - \frac{i \pi}{4} \Bigl)=s_2(\theta) \,,
\end{equation}
where the superscript index $0$ labels the physical massless particle and the index $+$ labels the chiral auxiliary roots.
Similarly, the kernels associated with the scattering between antichiral massless particles and antichiral auxiliary roots are 
\begin{equation}
K_{\downarrow}^{-0}(\theta)=K_{\downarrow}^{0-}(\theta)= \frac{1}{2 \pi i} \frac{d}{d \theta} \log \tanh \Bigl( -\frac{\theta}{2} + \frac{i \pi}{4} \Bigl)=s_2(\theta) \,,
\end{equation}
\begin{equation}
K_{\uparrow}^{0-}(\theta)=K_{\uparrow}^{-0}(\theta)= -\frac{1}{2 \pi i} \frac{d}{d \theta} \log \coth \Bigl( -\frac{\theta}{2} + \frac{i \pi}{4} \Bigl)=s_2(\theta) \,.
\end{equation}

\paragraph{Convolutions.}
We will use the following standard definition for the convolution
\begin{equation}
\bigl[f \star g \bigl] (\theta) \equiv \int_{-\infty}^{+\infty} d \tilde{\theta} f(\tilde{\theta}) g(\tilde{\theta}- \theta) \,.
\end{equation}

\subsection{Thermodynamic limit}

We can now take the thermodynamic limit
\begin{equation}
    R\to\infty,\qquad N\to\infty\,,\qquad \frac{N}{R}\quad\text{fixed}\,,
\end{equation}
where $N$ generically denotes the number of any type of excitations. As always, it is convenient to introduce densities $\rho(\theta)$ of particles and densities $\bar{\rho}(\theta)$ of holes, in terms of which we can write the thermodynamic limit of the (logarithm of the) Bethe--Yang equations in the various sectors.

\paragraph{Massive sector.}
The densities are
\begin{enumerate}
    \item $\rho_Q(\theta)$ for massive particles of quantum number $Q=1, \dots, k-1$ and $\theta \in \mathbb{R}$;
    \item $\rho^{(\alpha)}_{\downarrow}(\theta)$ for  auxiliary roots of massive type, with $\alpha = 1, 2$ and $\text{Im}(\theta)=0$;
    \item $\rho^{(\alpha)}_{\uparrow}(\theta)$ for each auxiliary root of massive type, with $\alpha = 1, 2$ and $\text{Im}(\theta)=i \pi$.
\end{enumerate}
Taking the thermodynamic limit of the logarithms of~\eqref{eq1:BAEq_fund_bound_states} and~\eqref{eq2:BAEq_fund_bound_states} we obtain\footnote{We use that in a relativistic theory $\frac{d}{d \theta}q_P=E_P$.}
\begin{equation}
\label{eq:constraint_rhoQ_densities}
\rho_P+ \bar{\rho}_P= \frac{R}{2 \pi} E_P + \sum_{Q=1}^{k-1} \rho_Q \star K^{QP} + \sum_{\alpha=1,2}  \left( \rho^{(\alpha)}_\uparrow \star K^{yP}_\uparrow -\rho^{(\alpha)}_\downarrow \star K^{yP}_\downarrow \right) \,,
\end{equation}
\begin{equation}
\label{eq:constraint_rhoalpha_densities_updown}
\rho^{(\alpha)}_{\bbf}(\theta)+ \bar{\rho}^{(\alpha)}_{\bbf}(\theta)= \sum_{Q=1}^{k-1}\bigl[ \rho_Q \star K_{\bbf}^{Qy} \bigl] (\theta) \,, \quad \bbf= \downarrow \, , \, \uparrow .
\end{equation}

\paragraph{Chiral/antichiral sector.}

Recalling that the superscript symbols $+$ and $-$ correspond to chiral and antichiral particles respectively, we have
\begin{enumerate} 
    \item $\rho_0^{(\pm, \dot{\alpha})}(\theta)$ for  massless particles with $\theta \in \mathbb{R}$ and $\dot{\alpha}=1, \dots, N_\circ$;
    \item $\rho_{\downarrow}^{(\pm, \alpha)}(\theta)$ for auxiliary roots of massless type with $\alpha=1,2$ and $\text{Im}(\theta)=0$;
    \item $\rho_{\uparrow}^{(\pm, \alpha)}(\theta)$ for auxiliary roots of massless type with $\alpha=1,2$ and $\text{Im}(\theta)= i \pi$.
\end{enumerate}
$N_\circ$ refers to the number of indices $\dot{\alpha}$ and while from the Bethe equations it seems this number should be $N_\circ=2$ comparisons with semiclassical results seem to suggest this number is actually $N_\circ=1$~\cite{Frolov:2023wji}. Here we will leave this number free.
The associated equations are obtained by taking the thermodynamic limit of the logarithms of~\eqref{eq:Bethe_chiral_rl_bs}, \eqref{eq:Bethe_auxiliary_chiral_rl_bs}, \eqref{eq:Bethe_antichiral_rl_bs} and~\eqref{eq:Bethe_auxiliary_antichiral_rl_bs}, and are given by
\begin{equation}
\label{eq:massless_densities_physical}
\rho_0^{(\pm)}(\theta)+ \bar{\rho}_0^{(\pm)}(\theta)= \frac{R}{2 \pi} E^{(\pm)} + N_\circ \bigl[ \rho_0^{(\pm)} \star K^{00}\bigl] (\theta)+ \sum_{\alpha=1,2} \bigl( \rho^{(\pm, \alpha)}_\downarrow - \rho^{(\pm, \alpha)}_\uparrow  \bigl) \star s_2  \,,
\end{equation}
\begin{equation}
\label{eq:massless_densities_auxiliary}
\rho^{(\pm, \alpha)}_{\bbf}(\theta)+ \bar{\rho}_{\bbf}^{(\pm, \alpha)}(\theta)=  N_\circ  \bigl[ \rho^{(\pm)}_0 \star s_2 \bigl] (\theta) \,, \quad \bbf= \downarrow \, , \, \uparrow \,,
\end{equation}
where we removed the index $\dot{\alpha}$ from the equations, which does not play any role and replaced $\sum_{\dot{\alpha}} \to N_\circ$, being $N_\circ$ the number of massless modes.

\subsection{Simplified equations for densities}

Each kernel for physical-physical interaction contain a minimal part and a CDD-type factor. Following~\cite{Fendley:1991ve, Fendley:1992dm, Bombardelli:2018jkj} we show how it is possible to make the minimal part disappear from the equations for the densities. 

\paragraph{Massive sector.}
Note first the identities
\begin{equation}
\label{eq:useful_integral_reltions}
\begin{split}
&K^{Py}_\uparrow(\theta_1-v) * K^{yQ}_\uparrow (v-\theta_2)=+\frac{1}{2} K_\sigma^{PQ}(\theta_{12}) + \langle P+Q \rangle_{\theta_{12}}\,,\\
&K^{P y}_\downarrow(\theta_1-v) * K^{y Q}_\downarrow (v-\theta_2)=+\frac{1}{2} K_\sigma^{PQ}(\theta_{12})\,,
\end{split}
\end{equation}
Using~\eqref{eq:constraint_rhoalpha_densities_updown}
we can then write
\begin{equation}
\rho^{(\alpha)}_{\downarrow}(\theta)= -\bar{\rho}_{\downarrow}^{(\alpha)}(\theta)+ \sum_{Q=1}^{k-1}\bigl[ \rho_Q \star K_{\downarrow}^{Qy} \bigl] (\theta) \,.
\end{equation}
Substituting this expression into~\eqref{eq:constraint_rhoQ_densities} we get
\begin{equation}
\begin{split}
\rho_P+ \bar{\rho}_P=& +\frac{R}{2 \pi} E_P + \sum_{Q=1}^{k-1} \rho_Q \star K^{QP} + \sum_{\alpha=1,2}  \rho^{(\alpha)}_\uparrow \star K^{yP}_{\uparrow} \\
& +  \sum_{\alpha=1,2}  \bar{\rho}^{(\alpha)}_\downarrow \star K^{yP}_{\downarrow}  -  \sum_{\alpha=1,2} \sum_{Q=1}^{k-1}\rho_Q \star K_\downarrow^{Qy} \star K_\downarrow^{yP} \,.
\end{split}
\end{equation}
Using now the identity in the second row of~\eqref{eq:useful_integral_reltions} we see that the minimal part of the physical-physical kernel cancels out and we end up with
\begin{equation}
\label{eq:constraint_rhoQ_densities2}
\begin{split}
&\rho_P+ \bar{\rho}_P= \frac{R}{2 \pi} E_P + \sum_{Q=1}^{k-1}\rho_Q \star K_\Phi^{QP} + \sum_{\alpha=1,2}  \left(\rho^{(\alpha)}_\uparrow \star K^{yP}_{\uparrow}   +\bar{\rho}^{(\alpha)}_\downarrow \star K^{yP}_{\downarrow}\right) \,,\\
&\rho_{\bbf}^{(\alpha)}+ \bar{\rho}_{\bbf}^{(\alpha)}= \sum_{Q=1}^{k-1} \rho_Q \star K_{\bbf}^{Qy}  \,, \quad \bbf= \downarrow \, , \, \uparrow .
\end{split}
\end{equation}

\paragraph{Massless sector.}
Repeating the same argument on the equations for massless particles~(\eqref{eq:massless_densities_physical}, \eqref{eq:massless_densities_auxiliary}) and using the fact that
\begin{equation}
\phi_0=s_2 *s_2\,,
\end{equation}
we obtain
\begin{equation}
\label{eq:massless_densities_simpl}
\begin{split}
&\rho_0^{(\pm)}(\theta)+ \bar{\rho}_0^{(\pm)}(\theta)= \frac{R}{2 \pi} E^{(\pm)} + N_\circ \bigl[ \rho_0^{(\pm)} \star s_2\bigl] (\theta)+ \sum_{\alpha=1,2} \bigl( \rho^{(\pm, \alpha)}_\downarrow + \bar{\rho}^{(\pm, \alpha)}_\uparrow  \bigl) \star s_2  \,,\\
&\rho^{(\pm, \alpha)}_{\bbf}(\theta)+ \bar{\rho}_{\bbf}^{(\pm, \alpha)}(\theta)=  N_\circ  \bigl[ \rho^{(\pm)}_0 \star s_2 \bigl] (\theta) \,, \quad \bbf= \downarrow \, , \, \uparrow \,.
\end{split}
\end{equation}

\subsection{Ground-state TBA equations}
\label{subsec:TBA_eq}

The free energy of the model under discussion can  be written in the form
\begin{equation}
\mathcal{F}=\mathcal{F}_{m}+\mathcal{F}_{+}+\mathcal{F}_{-} \,,
\end{equation}
where we highlighted the contribution of massive, chiral and antichiral particles, respectively. Their form and the result of imposing $\delta\mathcal{F}=0$ is worked out in appendix~\ref{app:derivation_TBA}, and for all intents and purposes we treat the massive, chiral and antichiral sectors as if they were decoupled theories. 
Since in the simplified equations for densities~\eqref{eq:constraint_rhoQ_densities2} and~\eqref{eq:massless_densities_simpl} the roles of some auxiliary densities $\rho$ and $\bar{\rho}$ are inverted it is then convenient to use an inverted convention for their Y functions in order to have TBA equations written in a standard form. We define
\begin{equation}
\begin{split}
&Y^{(\alpha)}_{\uparrow}= e^{i \gamma^{(\alpha)}_{\uparrow}} \frac{\bar{\rho}^{(\alpha)}_{\uparrow}}{\rho^{(\alpha)}_{\uparrow}}\, ,\ \qquad \qquad Y^{(\alpha)}_{\downarrow}=  e^{-i \gamma^{(\alpha)}_{\downarrow}}\frac{\rho^{(\alpha)}_{\downarrow}}{\bar{\rho}^{(\alpha)}_{\downarrow}}\,,\\
&Y_{\downarrow}^{(\pm, \alpha)}=e^{i \gamma^{(\pm, \alpha)}_{\downarrow}} \frac{\bar{\rho}^{(\pm, \alpha)}_{\downarrow}}{\rho^{(\pm, \alpha)}_{\downarrow}}\,,\qquad Y_{\uparrow}^{(\pm, \alpha)}=e^{-i \gamma^{(\pm, \alpha)}_{\uparrow}} \frac{\rho^{(\pm, \alpha)}_{\uparrow}}{\bar{\rho}^{(\pm, \alpha)}_{\uparrow}}\,.
\end{split}
\end{equation}
We use instead the standard definition for physical Y functions:
\begin{equation}
\label{eq:Y_as_ratio_of_rhos}
Y_P= \frac{\bar{\rho}_P}{\rho_P} \hspace{3mm},\hspace{3mm} Y^{(\pm)}_0= \frac{\bar{\rho}^{(\pm)}_0}{\rho^{(\pm)}_0} \,.
\end{equation}
The stationary condition for the free energy leads to the following set of TBA equations.
\paragraph{Massive sector.}
Requiring that $\delta \mathcal{F}_m=0$ we obtain
\begin{equation}
\label{eq:TBA_massive_phys}
\begin{split}
 \log Y_P= &\;\frac{E_P}{T} -i \frac{P}{k} \sum_{\alpha=1,2} \gamma^{(\alpha)}_\downarrow  -  \sum_{Q=1}^{k-1}   \log \bigl(1 +\frac{1}{Y_Q} \bigl) \star K_\Phi^{QP} \\
 &- \sum_{\alpha=1,2} \log \bigl(1 +\frac{e^{i \gamma^{(\alpha)}_\uparrow}}{Y^{(\alpha)}_\uparrow}\bigl) \star K_{\uparrow}^{yP}- \sum_{\alpha=1,2} \log \bigl(1 +\frac{e^{-i \gamma^{(\alpha)}_\downarrow}}{Y^{(\alpha)}_\downarrow}\bigl) \star K_{\downarrow}^{yP} \,,\\
 \log  Y^{(\alpha)}_\bbf = &-\sum_{Q=1}^{k-1}   \log \bigl(1 +\frac{1}{Y_Q} \bigl) \star K_\bbf^{Q y} \,.
 \end{split}
\end{equation}

\paragraph{Massless sector.}
The request that $\delta \mathcal{F}_{\pm}=0$ leads to
\begin{equation}
\label{eq:TBA_massless_Y_phys}
\begin{split}
 \log Y^{(\pm)}_0=&\;\frac{E^{(\pm)}_0}{T}  -\frac{i}{2}  \sum_{\alpha=1,2} \gamma^{(\pm, \alpha)}_\uparrow - N_\circ  \log \bigl(1 + \frac{1}{Y^{(\pm)}_0} \bigl) \star s_2 \\
 &- \sum_{\alpha=1,2}   \log \bigl(1 +\frac{e^{i \gamma^{(\pm, \alpha)}_{\downarrow}}}{Y_\downarrow^{(\pm, \alpha)}} \bigl) \star s_2 - \sum_{\alpha=1,2}   \log \bigl(1 +\frac{e^{-i \gamma^{(\pm, \alpha)}_{\uparrow}}}{Y_\uparrow^{(\pm, \alpha)}} \bigl) \star s_2 \,,\\
 \log Y_\bbf^{(\pm, \alpha)}  = &-N_\circ   \log \bigl(1 +\frac{1}{Y^{(\pm)}_0} \bigl) \star s_2 \, . 
 \end{split}
\end{equation}

\paragraph{Free energy.}
The free energy at the thermodynamic equilibrium is then given by
\begin{multline}
f=- T \sum_{P=1}^{k-1} \int_{-\infty}^{+\infty} \frac{d\theta}{2 \pi} \left(E_P -i \frac{T P}{k} \sum_{\alpha=1,2} \gamma^{(\alpha)}_\downarrow \right) \log \bigl( 1 + \frac{1}{Y_P} \bigl)\\
- N_\circ T \sum_{A= \pm} \int_{-\infty}^{+\infty} \frac{d\theta}{2 \pi} \left(E^{(A)}_0 -\frac{i}{2} T \sum_{\alpha=1,2} \gamma^{(A, \alpha)}_\uparrow \right) \log \bigl( 1 + \frac{1}{Y^{(A)}_0} \bigl) 
\end{multline}
where we labelled by $f \equiv \frac{\mathcal{F}}{R}$ the free energy per unit of volume.

From now on we consider the following simple twist used for example in \cite{Frolov:2009in,Frolov:2021bwp} 
\begin{equation}
\gamma^{(\alpha)}_a= (-1)^\alpha (\pi + \mu)\,, \quad \gamma^{(\pm, \alpha)}_a= (-1)^\alpha(\pi + \mu)\,, \qquad \alpha =1, 2\,.
\end{equation}
In string theory the case $\mu =0$ corresponds to the even-winding number sector with periodic fermions and supersymmetric vacuum. The value $\mu= \pi$ corresponds instead to the odd-winding number sector, with antiperiodic fermions and nonsupersymmetric vacuum.
Defining $\mu_\alpha=(-1)^\alpha \mu$ we have the following equations for massive and massless particles
\begin{equation}
\label{eq:TBA_massive_phys_simple_twist}
\begin{split}
 \log Y_P= &+\frac{1}{T} E_P  -  \sum_{Q=1}^{k-1}   \log \bigl(1 +\frac{1}{Y_Q} \bigl) \star K_\Phi^{QP} \\
 &- \sum_{\alpha=1,2} \log \bigl(1 -\frac{e^{i \mu_\alpha}}{Y^{(\alpha)}_\uparrow}\bigl) \star K_{\uparrow}^{yP}- \sum_{\alpha=1,2} \log \bigl(1 -\frac{e^{-i \mu_\alpha}}{Y^{(\alpha)}_\downarrow}\bigl) \star K_{\downarrow}^{yP} \,,\\
 \log  Y^{(\alpha)}_\bbf = &-\sum_{Q=1}^{k-1}   \log \bigl(1 +\frac{1}{Y_Q} \bigl) \star K_\bbf^{Q y} \,,
 \end{split}
\end{equation}
\begin{equation}
\begin{split}
 \log Y^{(\pm)}_0=+&\frac{1}{T} E^{(\pm)}_0 - N_\circ  \log \bigl(1 + \frac{1}{Y^{(\pm)}_0} \bigl) \star s_2 \\
 &- \sum_{\alpha=1,2}   \log \bigl(1 -\frac{e^{i \mu_\alpha}}{Y_\downarrow^{(\pm, \alpha)}} \bigl) \star s_2 - \sum_{\alpha=1,2}   \log \bigl(1 -\frac{e^{-i \mu_\alpha}}{Y_\uparrow^{(\pm, \alpha)}} \bigl) \star s_2 \,,\\
 \log Y_\bbf^{(\pm, \alpha)}  = &-N_\circ   \log \bigl(1 +\frac{1}{Y^{(\pm)}_0} \bigl) \star s_2 \, . 
 \end{split}
\end{equation}
with free-energy given by
\begin{equation}
\label{eq:free_energy_simpl_Y}
f=- T \sum_{P=1}^{k-1} \int_{-\infty}^{+\infty} \frac{d\theta}{2 \pi} E_P  \log \bigl( 1 + \frac{1}{Y_P} \bigl)- N_\circ T \sum_{A= \pm} \int_{-\infty}^{+\infty} \frac{d\theta}{2 \pi} E^{(A)}_0 \log \bigl( 1 + \frac{1}{Y^{(A)}_0} \bigl) \,.
\end{equation}

\subsection{Small-twist solutions}
For small twist $\mu \ll1$ we expect the free energy of the system to vanish since the vacuum state is supersymmetric. This requires $Y_P \to \infty$ and $Y^{(\pm)}_0 \to \infty$ in the limit $\mu \to 0$. From the TBA equations then it holds that both $Y^{(\alpha)}_a$ and $Y^{(\pm, \alpha)}_a$ must tend to $1$ in the limit.
To see that such a solution exists we can expand the Y functions for small values of $\mu$ following the argument in~\cite{Frolov:2009in, Frolov:2023wji}. Since the auxiliary functions tend to one in the limit of null twist we take the following expansion
\begin{equation}
Y^{(\alpha)}_a(\theta)=1+B^{(\alpha)}_a(\theta) \mu+ O(\mu^2)\,, \qquad Y^{(\pm, \alpha)}_a=1+B^{(\pm, \alpha)}_a(\theta) \mu+ O(\mu^2)\,.
\end{equation}
Now we replace the expressions above in the TBA equations for physical particles. Using that $Y_P$ and $Y^{(\pm)}_0$ diverge in the limit of small twist we obtain
\begin{equation}
\begin{split}
&\log Y_P= \frac{E_P}{T} - 2\log \mu * ( \langle 1 \rangle + \langle k-1 \rangle )+\mathcal{R}(\theta)\,,\\
&\log Y^{(\pm)}_0= \frac{E_0}{T} - 4\log \mu * s_2 +\mathcal{R}^{(\pm)}_0(\theta)
\end{split}
\end{equation}
where $\mathcal{R}(\theta)$ and $\mathcal{R}_0(\theta)$ are functions of $B^{(\alpha)}_a$ and $B^{(\pm, \alpha)}_a$ given by
\begin{equation}
\begin{split}
&\mathcal{R}=-\log \left( B^{(1)}_\uparrow +i \right) \left( B^{(2)}_\uparrow -i \right) * \langle P \rangle -\log \left( B^{(1)}_\downarrow -i \right) \left( B^{(2)}_\downarrow +i \right) * \langle k-P \rangle\,,\\
&\mathcal{R}^{(\pm)}_0=-\log \left( B^{(\pm, 1)}_\downarrow +i \right) \left( B^{(\pm, 2)}_\downarrow -i \right) * s_2 -\log \left( B^{(\pm, 1)}_\uparrow -i \right) \left( B^{(\pm, 2)}_\uparrow +i \right) * s_2\,.
\end{split}
\end{equation}
Using that
\begin{equation}
\label{eq:integral_angle_br}
\int^{+\infty}_{-\infty} dz \langle P \rangle _{z}=  \frac{k-P}{k}\,, \qquad \int^{+\infty}_{-\infty} dz \langle k-P \rangle _{z}=  \frac{P}{k} \,, \qquad  \int^{+\infty}_{-\infty} dz s_{2}(z)= \frac{1}{2} 
\end{equation}
we find
\begin{equation}
Y_P=\frac{e^{\frac{E_P}{T}}}{\mu^2} e^{\mathcal{R}(\theta)}\,, \qquad Y^{(\pm)}_0=\frac{e^{\frac{E^{(\pm)}_0}{T}}}{\mu^2} e^{\mathcal{R}_0(\theta)}\,.
\end{equation}
Plugging the relations above into the TBA equations for the auxiliary roots we find that it must hold
\begin{equation}
B^{(\alpha)}_a=B^{(\pm, \alpha)}_a=0 \implies \mathcal{R}=\mathcal{R}^{(\pm)}_0=0\,.
\end{equation}
The leading order of the Y functions at small twist is then
\begin{equation}
\begin{split}
&Y_P=\frac{e^{\frac{E_P}{T}}}{\mu^2} + O(\frac{1}{\mu})\,, \qquad Y^{(\pm)}_0=\frac{e^{\frac{E^{(\pm)}_0}{T}}}{\mu^2} + O(\frac{1}{\mu})\,,\\
&Y^{(\alpha)}_a(\theta)=1+ O(\mu^2)\,, \qquad Y^{(\pm, \alpha)}_a=1+ O(\mu^2)\,.
\end{split}
\end{equation}
The free energy in this limit is then given by
\begin{equation}
\label{eq:free_energy_simpl_Y2}
f=- T \mu^2 \sum_{P=1}^{k-1} \int_{-\infty}^{+\infty} \frac{d\theta}{2 \pi} E_P  e^{-E_P/T}- N_\circ T \mu^2 \sum_{A= \pm} \int_{-\infty}^{+\infty} \frac{d\theta}{2 \pi} E^{(A)}_0 e^{-E^{(A)}_0/T} \,.
\end{equation}
This is consistent with the fact that in the limit $\mu \to 0$ (when supersymmetry is restored) the vacuum energy vanishes.

\subsection{Y system}

Let us now work out the Y system by making use of the right-inverse $s_{k}^{-1}$ of the Cauchy kernel~\eqref{eq:Cauchy_kernel_definition}.

\paragraph{Massive sector.}
The TBA equations for massive particles involve sums over $k-1$ terms, associated with the massive bound states.
These sums can be removed and the equations can be simplified as explained appendix~\ref{app:simplification_TBA}. For physical particles we get
\begin{equation}
\label{eq:massive_simplified_exp}
\begin{split}
&Y_P(\theta- \frac{i \pi}{k}) Y_P(\theta+ \frac{i \pi}{k})=\frac{\left(1+Y_{P+1}(\theta) \right) \left( 1+Y_{P-1}(\theta) \right)}{\prod_{\alpha=1,2} \left(1-\frac{e^{i \mu_\alpha}}{Y^{(\alpha)}_\uparrow} \right)^{\delta_{P 1}} \left(1-\frac{e^{-i \mu_\alpha}}{Y^{(\alpha)}_\downarrow} \right)^{\delta_{P k-1}} } \,, \qquad P=1, \dots, k-1\,,\\
&Y_0=Y_k=0\,.
\end{split}
\end{equation}
For auxiliary particles we obtain
\begin{equation}
\label{eq:aux_Y_simplified_new}
\begin{split}
& Y^{(\alpha)}_\uparrow(\theta+ \frac{i \pi}{2}) Y^{(\beta)}_\downarrow(\theta- \frac{i \pi}{2}) =\prod^{k-1}_{Q=1} \left(1+\frac{1}{Y_{Q}(\theta+\frac{i \pi}{2} - \frac{i \pi}{k} Q)} \right)^{-1} \,,\\
& Y^{(\alpha)}_\uparrow(\theta- \frac{i \pi}{2}) Y^{(\beta)}_\downarrow(\theta+ \frac{i \pi}{2}) =\prod^{k-1}_{Q=1} \left(1+\frac{1}{Y_{Q}(\theta-\frac{i \pi}{2} + \frac{i \pi}{k} Q)} \right)^{-1} \,.
\end{split}
\end{equation}
The Y-system for auxiliary particles does not depend on the indices $\alpha$ and $\beta$, which take values $\in \{1, 2\}$. 
Remark that up to sending $Y \to {1}/{Y}$ and doubling the auxiliary excitations (through the index $\alpha=1,2$) this Y system is identical to the one obtained in~\cite{Fabbri:2013wua}
associated to relevant deformations of $(\mathbb{C} \mathbb{P}^{k-1} )_p \times U(1)$ conformal cosets  --- specifically, for the case $p=2$.%
\footnote{In particular, our equations~\eqref{eq:massive_simplified_exp} and~\eqref{eq:aux_Y_simplified_new} are closely related to equations (3.7) and (3.5) of~\cite{Fabbri:2013wua} when $p=2$. This similarity is very helpful for simplifying the TBA equations. We thank Roberto Tateo for pointing this out to us.}

\paragraph{Massless sector.}
The TBA equations for massless particles are simplified by using the crossing equations, together with the fact that 
\begin{equation}
E^{(\pm)}_0(\theta- \frac{i \pi}{2})+E^{(\pm)}_0(\theta+ \frac{i \pi}{2})=0\,.
\end{equation}
In this case we obtain
\begin{equation}
\begin{split}
 &Y^{(\pm)}_0(\theta+\frac{i \pi}{2}) Y^{(\pm)}_0(\theta+\frac{i \pi}{2})=\left(1 +\frac{1}{Y^{(\pm)}_0} \right)^{-N_\circ}   \prod_{\alpha=1,2}  \left(1 -\frac{e^{i \mu_\alpha}}{Y_\downarrow^{(\pm, \alpha)}} \right)^{-1} \left(1 -\frac{e^{-i \mu_\alpha}}{Y_\uparrow^{(\pm, \alpha)}} \right)^{-1} \,,\\
 &Y_\bbf^{(\pm, \alpha)}(\theta+ \frac{i \pi}{2})Y_\bbf^{(\pm, \alpha)}(\theta- \frac{i \pi}{2})  = \left(1 +\frac{1}{Y^{(\pm)}_0} \right)^{-N_\circ} \, . 
 \end{split}
\end{equation}

\section{Solutions at \texorpdfstring{$\mu= \pi$}{mu=pi} and small-volume limit} 
\label{sec:central_charge}

After identifying $T=\frac{1}{L}$, it is possible to compute the $L\to0$ limit of the ground state energy by the ``dilogarithm trick'', see appendix~\ref{sec:app_Rogers}. If this non-conformal theory originates from a UV CFT, we expect that as $L\to0$ the ground-state energy diverges as $1/L$, and that we may read off the central charge as
\begin{equation}
    \mathcal{E}_{\text{CFT}}=-\frac{\pi}{6}\frac{c}{L}\,.
\end{equation}
To compute this at small~$L$ it is sufficient to find constant solutions to the Y system at the twist value $\mu=\pi$. In the massive sector%
\footnote{The incidence matrix $I_{P a}$ is given by $I_{P \bbf} \equiv \delta_{P,1} \delta_{\bbf, \uparrow}+\delta_{P, k-1} \delta_{\bbf, \downarrow}$, see~\eqref{IPQ_IPbullet_definitions}.}
\begin{equation}
\label{eq:Y_syst_constant}
\begin{split}
&Y^2_P =\frac{\left(1+Y_{P+1} \right) \left( 1+Y_{P-1} \right)}{\prod_{a=\uparrow, \downarrow} \left(1+\frac{1}{Y_a} \right)^{Z_\alpha I_{Pa}} }\,, \qquad Y_0=Y_k=0\,,\\
& Y_\uparrow Y_\downarrow =\prod^{k-1}_{Q=1} \left(1+\frac{1}{Y_{Q}} \right)^{-1} \,,\\
\end{split}
\end{equation}
and in the massless sector
\begin{equation}
\label{eq:Y_syst_constant_massless}
\begin{split}
 &(Y^{(\pm)}_0 )^2=  {\left(1 +\frac{1}{Y^{(\pm)}_0} \right)^{-N_\circ}}  \left(1 +\frac{1}{Y_\downarrow^{(\pm)}} \right)^{-Z_\alpha} \left(1 +\frac{1}{Y_\uparrow^{(\pm)}} \right)^{-Z_\alpha} \,,\\
 &(Y_\bbf^{(\pm)})^2  = \left(1 +\frac{1}{Y^{(\pm)}_0} \right)^{-N_\circ} \, .
 \end{split}
\end{equation}
Since at this value of the twist parameter there is no distinction between $\alpha=1$ and $\alpha=2$ in the TBA equation we omitted to keep track of the index $\alpha$ and replaced the sums $\sum_{\alpha=1,2} \to Z_\alpha$, where $Z_\alpha$ corresponds to the number of auxiliary roots --- which in turns is related to the amount of supersymmetry. Keeping this number implicit allows us to discuss both the case of refs.~\cite{Fendley:1991ve, Fendley:1992dm} ($Z_\alpha=1$, with $\mathcal{N}=(2,2)$ susy) and our case ($Z_\alpha=2$, with twice as many supercharges).
Similarly, we leave the number of massless modes $N_\circ$ implicit for now. Clearly the central charge depends on the total ground-state energy. However, since our system is decoupled it is interesting to separately consider the massive and massless contributions to $\mathcal{E}$ and hence to~$c$.

\subsection{Central charges}

In the relativistic limit of~\cite{Frolov:2023lwd} which we are considering here, the massive and massless modes decouple already at the level of the S~matrix.
This is these different kinematic regimes have different scaling as we approach the relativistic regime. The decoupling is not entirely surprising if we think that a massless relativistic particle can be formally obtained from a massive one by means of an infinite boost. Boosting one of two particles in the S~matrix is the same as sending $\theta_{12}\equiv\theta_1-\theta_2\to\pm\infty$, and in this limit the scattering is expected to trivialise.
This decoupling simplification allows us to study the massive and massless sectors independently and write the associated decoupled Y-systems. We stress that this is not the case in the original non-relativistic theory, where massive and massless particles interact with each other, and massless modes are related to $k$-particle massive bound states~\cite{Frolov:2025uwz} and the massive and massless excitations are coupled in the TBA~\cite{Frolov:2025tda}. 
In this section, we will show a further decoupling of certain Y functions in the UV limit of the theory.

We start considering the Y system~\eqref{eq:Y_syst_constant} for massive particles. 
Note that the system of equations is invariant under exchanging $\{Y_\uparrow \leftrightarrow Y_\downarrow, Y_P \leftrightarrow Y_{k-P} \}$ and the solutions must respect the same symmetry. We expect this system to have only one physical solution (with real energy). If this is the case,%
\footnote{We have checked that this holds for a few values of~$k$.}
then solutions must have the form 
\begin{equation}
\label{eq:sym_solutions_const_Y}
Y_P=Y_{k-P}\,, \qquad Y_\uparrow=Y_\downarrow \equiv Y_{\text{aux}}\,.
\end{equation}
Much like it was observed in~\cite{Fendley:1992dm} we find that with this assumptions it is possible to find a physical, closed-form solution for any~$k=2,3,4,\dots$.

\paragraph{Massive sector,  $Z_\alpha=1$.}
If we take $Z_\alpha=1$ then this system is identical to the one studied in section 4 of~\cite{Fendley:1992dm} and admits the following nontrivial solution
\begin{equation}
\label{eq:FI_Y_solution}
Y_P=\frac{\sin \left( \frac{(2P+3) \pi}{2(k+1)} \right) \sin \left( \frac{(2P-1) \pi}{2(k+1)} \right)}{\sin^2 \left( \frac{\pi}{k+1} \right) }\,, \qquad Y_\uparrow=Y_\downarrow= \frac{\sin \frac{\pi}{2 (k+1)}}{\sin \frac{3\pi}{2 (k+1)}}\,,
\end{equation}
respecting the symmetry~\eqref{eq:sym_solutions_const_Y}.
Using the formulae in appendix~\ref{sec:app_Rogers} and in particular eq.~\eqref{eq:app_gr_state_energy_UV} after identifying the temperature $T=1/L$ we find the ground state energy
\begin{equation}
\begin{split}
L \mathcal{E}=&+\frac{1}{2\pi} \sum^{k-1}_{P=1}   \int^{0}_{1/Y_P} d z \ \left( \frac{\log \left(1 + z \right)}{z} - \frac{\log z}{1 + z} \right)\\
&+\frac{1}{2\pi} 2  \int^{1}_{1/Y_{\text{aux}}} d z_k \ \left( \frac{\log \left(1 + z_k \right)}{z_k} - \frac{\log z_k}{1 + z_k} \right) \,.
\end{split}
\end{equation}
The factor $2$ in the second line comes from the fact that there are two auxiliary roots, $\uparrow$ and $\downarrow$. Performing the integrals we obtain the following UV central charge
\begin{equation}
c= -\frac{6}{\pi} L \mathcal{E}=3 \frac{k-1}{k+1}\,,\qquad k=2,3,4,\dots\,.
\end{equation}
This result was already known from~\cite{Fendley:1992dm} and corresponds to the central charge of the $A_{k-1}$ series of $\mathcal{N}=(2,2)$ minimal models. This is in agreement with the expectations; indeed TBA equations for $Z_\alpha=1$ correspond to a deformation of this class of models by their most relevant operator~\cite{Fendley:1992dm}.

\paragraph{Massive sector,  $Z_\alpha=2$.}
Let us now focus on $Z_\alpha=2$, which is the case of interest for the model emerging from the relativistic limit of AdS$_3$ superstrings. In this case, we obtain that the only physical solution to the Y system~\eqref{eq:Y_syst_constant} corresponds to
\begin{equation}
\label{eq:sol_Y_massive_Z_2_aux}
\begin{split}
&Y_\uparrow=Y_\downarrow=0\,,\\
&Y_{P}=\frac{\sin \left( \frac{\pi}{k} (P-1) \right) \sin \left( \frac{\pi}{k} (P+1) \right)}{\sin^2 \frac{\pi}{k}}\,, \qquad P=1, \, , \dots, \, k-1\,.
\end{split}
\end{equation}
Note that in this solution the following Y-functions vanish:
\begin{equation}
\label{eq:theyvanish}
    Y_\uparrow=Y_\downarrow=0\,,\qquad Y_1=Y_{k-1}=0\,.
\end{equation}
The remaining Y-functions obey the Y-system
\begin{equation}
\label{eq:sol_Y_massive_Z_2_phys}
\begin{split}
&Y^2_2=1+Y_3\,,\\
&Y_3^2=(1+Y_2)(1+Y_4)\,, \quad \dots \quad Y_{k-3}^2=(1+Y_{k-4})(1+Y_{k-2})\,,\\
&Y^2_{k-2}=1+Y_{k-3}\,,
\end{split}
\end{equation}
which is indeed solved by~\eqref{eq:sol_Y_massive_Z_2_aux}.
This suggests that in the $L\to0$ limit, the Y-functions~\eqref{eq:theyvanish} decouple (and behave like in a free model). For $k > 3$ the remaining, non-trivial Y-functions are those of minimal S-matrices associated with $A_{k-3}$ Dynkin diagrams in the classification of~\cite{Ravanini:1992fi}.
The solutions for the Y functions for these theories can be found for example in~\cite{Ravanini:1992fi, Klassen:1989ui} and the contribution to the central charge is given by $2-\frac{6}{k}$, corresponding to $\mathbb{Z}_{k-2}$ parafermions~\cite{Fateev:1985mm}.

In total, we can compute the central charge and find  
\begin{equation}
\label{eq:central_charge_k_free}
c=-\frac{6}{\pi} L \mathcal{E} = 6 \left(1-\frac{1}{k} \right) \,.
\end{equation}
As far as we can tell, this central charge does not correspond to that of a $\mathcal{N}=(2,2)$ or $\mathcal{N}=(4,4)$ minimal model. Recall that our model has the same amount of supersymmetry as a  $\mathcal{N}=(4,4)$ theory but the supercharges obey a different algebra --- two copies of the $\mathcal{N}=(2,2)$  supersymmetry algebra, with the same central elements. This, and the fact that the Y-system decouples in two pieces as $L\to0$, is suggestive that we may be dealing with two (or more) decoupled UV models. It is very suggestive to conjecture that one may be a free theory with $c=3$ and the other an $\mathcal{N}=(2,2)$ theory with $c=3-6/k$. However, the above discussion seems to instead suggest a different split, at least for $k\geq3$: a $c=4$ theory and a theory of $\mathbb{Z}_{k-2}$  parafermions with $c=2-6/k$. Such a non-supersymmetric splitting may indicate that our supersymmetry algebra does not lead to a good superconformal theory as $L\to0$.%
\footnote{The value of the central charge (especially the term $6/k$) is also reminiscent of $\mathcal{SW}$-algebras in the presence of NSNS flux (see, e.g., \cite{Fiset:2021azq,delaOssa:2024cgo}). We thank Enrico Marchetto for pointing this out to us.}

Finally, let us contrast our result for the central charge with that of Fendley and Intriligator~\cite{Fendley:1992dm}. Since we have twice as many supercharges as they do, the representations which we scatter are four-dimensional (rather than two-dimensional) and as such one might have expected that our central charge should be twice that of Fendley and Intriligator, but ours is actually smaller:
\begin{equation}
    c<2\,c_{\text{FI}}=6-\frac{4}{k+1}\,.
\end{equation}
This is because while we doubled the number of particles of the model, we have done so by adding interactions (obtaining the representations by taking the tensor product, not the direct sum). Indeed, the structure of the TBA highlights that our degrees of freedom are not truly twice as many than those of~\cite{Fendley:1992dm}: we have the same momentum-carrying Y-functions, and we doubled the auxiliary ones.

\paragraph{Periodicity of the massive Y system.} By giving certain initial conditions to the Y functions at $\theta$ and $\theta+\frac{i \pi}{k}$, it is then possible to generate all the remaining Y functions at $\theta+\frac{2i \pi}{k}$, $\theta+\frac{3i \pi}{k}$, $\dots$ by applying the equations of the Y system iteratively. By doing so we observe that for $Z_\alpha=1$ the Y system is periodic and satisfies
\begin{equation}
Y(\theta)=Y(\theta+ i \pi P_k)\,, \qquad P_k=\frac{2}{k} (k+1)\,.
\end{equation}
From this period we can read out the conformal dimension of the operator perturbing the UV CFT~\cite{Zamolodchikov:1991et}
\begin{equation}
\Delta_k=1 -\frac{1}{P_k}= \frac{2+k}{2+2k}\,,
\end{equation}
in agreement with the results early obtained in~\cite{Fabbri:2013wua} for the special case $p=2$ of that paper.

On the opposite for $Z_\alpha=2$ the Y system turns out to not be periodic. Instead, no matter the starting initial conditions we choose for the Y functions, we observe that after a few iterations the functions $Y_{1}(\theta+i t)$, $Y_{k-1}(\theta+i t)$, $Y_{\uparrow}(\theta+i t)$ and $Y_{\downarrow}(\theta+i t)$ 
tend to zero exponentially fast for $\theta$ real and $t\to\pm\infty$. The remaining Y functions converge exponentially, for $|t|\gg1$ to a solution with $2\pi i$ periodicity.
Such a periodicity is the same as in the $A_{k-3}$ series of minimal S-matrices~\cite{Zamolodchikov:1991et,Ravanini:1992fi}. This is in agreement with our solution~\eqref{eq:sol_Y_massive_Z_2_aux} for constant Y functions.

\paragraph{UV decoupling and comparison with literature.}
The decoupling of $Y_1$, $Y_{k-1}$ and $Y^{(\a)}_{a}$ from the other Y functions in the UV limit of the TBA is far less obvious than the kinematical decoupling of massless and massive particles (which is already evident in the S~matrix). We do not have a first-principle explanation for this decoupling, but may be helpful to compare with a similar result observed in the past in the case $k=2$.
The case $k=2$ is special since for this value there are no bound states and all the auxiliary Y functions are connected to a single massive node (as $Y_1$ and $Y_{k-1}$ coincide).
In this particular case, the UV central charge~\eqref{eq:central_charge_k_free} turns out to be
\begin{equation}
c=3\,.
\end{equation}
This was already observed in~\cite{Fontanella:2019ury}; in that paper, the authors interpreted this central charge as a free theory limit containing two massive bosons and two massive fermions, which matches the content of the supersymmetric multiplet for this model. The authors suggested then to multiply this factor by two to take into account both the left and right representations of the original string theory, with the interpretation that in the UV limit the model reduces to a free theory for the original particles of the string theory. Note however that, as soon as we consider $k=3, 4,5, \dots$ the TBA and its UV limit are quite different from those of~\cite{Fontanella:2019ury}.%
\footnote{In~\cite{Fontanella:2019ury}, only the case $k=2$ was discussed, since crossing was only solved somewhat implicitly for $k=3,4,\dots$. Moreover, at $k=2$ they also interpreted ``our'' $m=1$ excitations as if they were massless. This is not relevant in the UV limit, where ours result agree with theirs.}
A numerical study of the $k=2$ TBA equations was performed in~\cite{Bombardelli:2018jkj},%
\footnote{Even if the relativistic S~matrices of~\cite{Fontanella:2019ury} and~\cite{Bombardelli:2018jkj} arise from two different limiting procedures, they coincide.}
which confirms $Y_1=Y_{\uparrow}=Y_\downarrow=0$  in the UV (see discussion around equation (5.50) of that paper). 
It would be interesting to implement a similar numerical study of the TBA equations for other values of $k$, in such a way as to provide the values of the $Y$ functions and the effective central charge away from the point $L =0$. This would provide a deeper insight into the reason why we observe this decoupling of the $Y$ functions at the UV point. 
Another parallel direction to be numerically addressed would be to find the excited states of the model.

\paragraph{Massless sector.}
In a similar way, we solve the massless-sector Y-system~\eqref{eq:Y_syst_constant_massless} when $Z_\alpha=2$.
We expect that the result will depend on the number of massless momentum carrying-modes~$N_\circ$. Nonetheless, we find that it is always
\begin{equation}
\label{eq:zero_soultion_Y_massless}
Y^{(\pm)}_{\uparrow}=Y^{(\pm)}_{\downarrow}=Y^{(\pm)}_0=0\,.
\end{equation}
This is the same result that we would have in a free theory, with the various Y-functions counted with their multiplicity. We find
\begin{equation}
c=N_\circ+2\,.
\end{equation}
So, in the case where $N_\circ=1$, we find what looks like a free supersymmetric theory with~$c=3$ in the UV. In the case $N_\circ=2$ we find $c=4$; this looks less like a free $\mathcal{N}=(2,2)$ theory, however it is not clear that this should be the case, looking back at our discussion for the massive sector.

It is also worth briefly comparing our results with those of~\cite{Bombardelli:2018jkj, Fontanella:2019ury}. The system which they considered is related to our case $N_\circ=1$. Indeed, they also had found $c=3$. It is not clear to us their argument to double this and obtain $c=6$ (they appear to disregard the mutual interaction of the massless momentum-carrying modes).

\paragraph{On the finite-size spectrum of the model.}
Owing to the decoupling of the massive and massless S~matrix, it is possible to study the relative finite-volume corrections separately. A complete study of such corrections would require the numerical solution of the TBA, which we postpone to a later work. Some general considerations can be already formulated based on the leading L\"uscher corrections, see e.g.~\cite{Janik:2010kd} for a review. In the massive TBA, we expect the finite-volume corrections to be exponentially suppressed by the mass scale, as $e^{-mL}$, while the Bethe--Yang equations provide a good approximation of the spectrum. Conversely, for the massless excitation wrapping is expected to contribute already at order $1/L$, i.e.\ at the same order of the Bethe--Yang equations.

\section{Conclusions and outlook}
\label{sec:conclusions}
In this paper we have worked out the TBA and Y-system for the relativistic limit constructed in~\cite{Frolov:2023lwd}. The model is strongly reminiscent of the $\mathcal{N}=(2,2)$ supersymmetric theory studied by Fendley and Intriligator~\cite{Fendley:1991ve, Fendley:1992dm}, but it has twice the amount of supersymmetry. Yet, the model does not have $\mathcal{N}=(4,4)$ SUSY; rather, its symmetries form two copies of the $\mathcal{N}=(2,2)$ algebra with the same bosonic charges. This is a natural structure from the point of view of the worldsheet string model in lightcone gauge~\cite{Lloyd:2014bsa} (in fact, very similar to what happens also for $AdS_5\times S^5$, see e.g.~\cite{Arutyunov:2009ga}), but an unusual one from the point of view of a two-dimensional relativistic theory.
To our knowledge, such relativistic theories have not been considered before. In this FPS model, the dynamics of the massive and massless modes decouple --- something that was  already clear from the study of their S~matrix~\cite{Frolov:2023lwd} --- and the TBA can be worked out. The massive modes yield a Y-system that can be cast in local form only after introducing different shifts for the physical and auxiliary Y functions, as early observed in~\cite{Fabbri:2013wua} for a similar class of models.
In the UV limit, where we derive the central charge, our system behaves quite differently from~\cite{Fendley:1992dm} and it seems related to a model of $\mathbb{Z}_{k-2}$ parafermions. It would be interesting to work out in detail whether and how this is the case.

It is not immediately clear whether the TBA which we have constructed has some implications for the ongoing challenge of understanding the spectrum of superstrings on mixed-flux AdS$_3\times $S$^3\times$T$^4$ which was our starting point, even if when restricting to some appropriate limit of that spectrum. Not only the relativistic limit requires $h\to0$, but it also requires focusing on a small sector of low-lying states. This is no problem at the level of the S~matrix, but problematic in the TBA since --- by its very construction --- the latter requires summing over all states of a theory. To make matters even more complicated, the original worldsheet model was non-relativistic, so that we expect a very substantial difference when computing the finite-volume or finite-energy partition function~\cite{Arutyunov:2007tc} --- which is not the case \textit{after} the limit.
Hence, for the purpose of understanding the spectrum, we believe it is most appropriate to consider the FPS  model as being defined in its own right, forgetting about its construction as a limit of a stringy construction. It is still an extremely interesting model, and well deserving of further study.
Similarly, it would also be very interesting to extend this analysis to the case of AdS$_3\times $S$^3\times$S$^3\times$S$^1$~\cite{Borsato:2015mma}, see also~\cite{Seibold:2025fnu} and references therein. There the relativistic limit would be more subtle, as we would be dealing with particles of mass $\alpha$ and $1-\alpha$, where $0<\alpha<1$ is a parameter encoding the ratio of the spheres' radii. Interestingly, as the lightcone gauge breaks all but one quarter of the supersymmetry, that limit is quite likely to yield a family of $\mathcal{N}=(2,2)$ theories, though it  is not clear if they would be the same as in~\cite{Fendley:1991ve, Fendley:1992dm}. We hope to return to some of these questions in the future.

\section*{Acknowledgments}
XK is grateful to the Department of Physics at the University of Padova for their hospitality. XK also acknowledges the generous support of the Fondation du Domaine de Villette for funding his research activities during the summer of 2023.
DP and AS are grateful to Matheus Fabri, Sergey Frolov, Enrico Marchetto, Roberto Tateo and Roberto Volpato for useful discussions, and to Alessandro Torrielli for correspondence. DP and AS expecially thank Roberto Tateo for enlightening discussions on the simplification of the Y~system. 
DP and AS thank the participants of the ``Integrability in low supersymmetry theories'' school \& workshop in Trani, Italy, for stimulating discussions.
DP and AS would like to acknowledge support from the European Union – NextGenerationEU, from the program STARS@UNIPD, under the project “Exact-Holography – A new exact approach to
holography: harnessing the power of string theory, conformal field theory, and integrable models” and
from the PRIN Project n. 2022ABPBEY, ``Understanding quantum field theory through its deformations''.
This work has received funding from the Deutsche Forschungsgemeinschaft (DFG, German Research Foundation) -- SFB-Gesch\"aftszeichen 1624 -- Projektnummer 506632645.
 A.S.~also acknowledges support from the CARIPLO Foundation under grant n.~2022-1886, and from the CARIPARO Foundation Grant under grant n.~68079.

\appendix

\section{Sketch of the derivation of the Bethe--Yang equations} 
\label{app:first_appendix}
Let us consider the relativistic S~matrix of~\cite{Frolov:2023lwd}. In this appendix we sketch the derivation of the Bethe--Yang equations, without reference to those of the full model~\cite{Seibold:2022mgg}. For simplicity, we will sketch the derivation using the nested coordinate Bethe Ansatz~\cite{Borsato:2012ss}, see also~\cite{Sfondrini:2014via}, rather than the algebraic construction of~\cite{Seibold:2022mgg}, but of course this is irrelevant for the final result.

We shall focus on the $4\times4$ blocks which make up the full S-matrix, as our goal is to find the form of the auxiliary scattering matrices. Let us first consider a level-I state, in the notation of~\cite{Sfondrini:2014via}. This is constructed out of a collection of states which scatter elastically among each other, \textit{i.e.}, by highest-weight states of a given representation, which we denote by~$\mathcal{X}_p^{\alpha}$, where $\alpha$ denotes the flavour and $p$ the momentum. Such a state is then taken as the ground state~$|0\rangle^{\text{II}}$ in the nesting procedure:
\begin{equation}
    \ket{0}^{\text{II}} = 
    \ket{\mathcal{X}_{p_1}^{\alpha_1}\dots \mathcal{X}_{p_M}^{\alpha_M}}^{\text{I}}\,.
\end{equation}
An excitation~$\mathcal{Y}(y)$, obtaining by acting with a lowering operator over this ``vacuum'', is of the form
\begin{equation}
    \ket{\mathcal{Y}(y)}^{\text{II}}=
    \sum_{j=1}^M \chi_j(y;\{p\})
    \ket{\mathcal{X}_{p_1}^{\alpha_1}\dots 
    \mathcal{Y}_{p_j}^{\alpha_j}
    \dots\mathcal{X}_{p_M}^{\alpha_M}}\,,
\end{equation}
where
\begin{equation}
    \chi_j(y;\{p\})=
    F_{\alpha_j}(y,p_j)\prod_{k=1}^{j-1} S^{\text{II,I}}_{\alpha_j\alpha_k}(y,p_j)\,.
\end{equation}
In practice, for $M=2$ excitations, which is all we need, if we denote the (Bosonic) highest-weight state by $\phi$ and the lowest-weight state by $\psi$,%
\footnote{The case of Fermionic highest-weight states can be dealt with similarly, it just results in some minus signs.}
\begin{equation}
    \ket{\mathcal{Y}(y)}^{\text{II}}=
    F_{\alpha_1}(y,p_1)\ket{\psi^{\alpha_1}_{p_1} \phi^{\alpha_2}_{p_2}}+
    F_{\alpha_2}(y,p_2) S^{\text{II,I}}_{\alpha_1\alpha_2}(y,p_1)\,\ket{\phi^{\alpha_1}_{p_1} \psi^{\alpha_2}_{p_2}}.
\end{equation}
Compatibility with scattering means that
\begin{equation}
\label{eq:scatteringcond}
    \mathbf{S}_{\pi} \ket{\mathcal{Y}(y)}^{\text{II}}
    = \ket{\mathcal{Y}(y)}^{\text{II}}_{\pi}\,,
\end{equation}
where the permuted wave-function is, for $M=2$,
\begin{equation}
    \ket{\mathcal{Y}(y)}^{\text{II}}_{\pi}=
    F_{\alpha_2}(y,p_2)\ket{\psi^{\alpha_2}_{p_2} \phi^{\alpha_1}_{p_1}}+
    F_{\alpha_1}(y,p_1) S^{\text{II,I}}_{\alpha_2\alpha_1}(y,p_2)\, \,\ket{\phi^{\alpha_2}_{p_2} \psi^{\alpha_1}_{p_1}}.
\end{equation}
We now use that the $4\times4$ scattering matrix has a XXZ form, with six non-vanishing matrix elements $\mathsf{A}^{\alpha\beta}(p_1,p_2)$, \dots, $\mathsf{F}^{\alpha\beta}(p_1,p_2)$. The Greek labels denote the block (\textit{e.g.}, massive, massless, \textit{etc.}), and the explicit form of the matrix elements can be found in appendix~B of~\cite{Frolov:2023lwd}.
Explicitly, compatibility with scattering~\eqref{eq:scatteringcond} results in two equations
\begin{equation}
\begin{aligned}
    S^{\text{II,I}}_{\alpha\beta}(y,p_2) F_{\alpha}(y,p_1)\mathsf{A}^{\alpha\beta}_{p_1p_2}=
    F_{\alpha}(y,p_1)\mathsf{D}^{\alpha\beta}_{p_1p_2}+
    S^{\text{II,I}}_{\alpha\beta}(y,p_1) F_{\beta}(y,p_2)
     \mathsf{C}^{\alpha\beta}_{p_1p_2}\,,\\
    F_{\beta}(y,p_2)\mathsf{A}^{\alpha\beta}_{p_1p_2}=
    F_{\alpha}(y,p_1)\mathsf{E}^{\alpha\beta}_{p_1p_2}+
    S^{\text{II,I}}_{\alpha\beta}(y,p_1) F_{\beta}(y,p_2)
     \mathsf{B}^{\alpha\beta}_{p_1p_2}\,,
\end{aligned}
\end{equation}
We can solve these equations in the case where both particles are massive, $m_i\neq 0$~mod$k$. 
It is then easy to check that a solution is
\begin{equation}
    F_{m}(\eta,\theta)= \frac{\mu(m)\,e^{\frac{m\pi i}{k}}e^{\tfrac{1}{2}\theta}}{e^{\theta}-e^{\frac{m\pi i}{k}}e^{\eta}}\,,\qquad
    S^{\text{II,I}}(\eta,\theta)=\frac{\cosh(\frac{\eta-\theta}{2}-\frac{m \pi i}{2k})}{\cosh(\frac{\eta-\theta}{2}+\frac{m \pi i}{2k})}\,,
\end{equation}
where we parametrised the $y,p$ by rapidities.
This is precisely what we find as the limit of the Bethe--Yang equations of~\cite{Seibold:2022mgg}.
Clearly, had we considered different values of $\mathsf{A}^{\alpha\beta}(p_1,p_2)$, \dots, $\mathsf{F}^{\alpha\beta}(p_1,p_2)$, corresponding \textit{e.g.} to massless representations, we would obtain different auxiliary S~matrices.

\section{Bound states and fusion}
\label{app:BS}
A necessary step in determining the TBA equations is to determine what configurations of particles dominate in the thermodynamic limit --- the Bethe strings. A pedagogical explanation of this topic can be found in~\cite{vanTongeren:2013gva}. We will show that for the model under discussion, the Bethe strings are composed of up to $k-1$ particles and correspond to the bound states of the model found in~\cite{Frolov:2023lwd}.

We start by recalling the fusions properties of the relativistic S-matrices in~\eqref{eq:aux_phys_dressing_m}, \eqref{minimal_sigma_ratio_of_R_functions} and~\eqref{general_formula_for_CDD_factor}, which satisfy
\begin{equation}
\label{eq:fusion_prop_dressing_phases}
\begin{aligned}
&\sigma(m_1,m_3;\theta+i \frac{\pi}{k}m_2)^{-2} \sigma(m_2,m_3;\theta-i \frac{\pi}{k}m_1)^{-2} =\sigma(m_1+m_2,m_3;\theta)^{-2} \,,\\
&\Phi(m_1,m_3;\theta+i \frac{\pi}{k}m_2) \Phi(m_2,m_3;\theta-i \frac{\pi}{k}m_1) =\Phi(m_1+m_2,m_3;\theta)
\end{aligned}
\end{equation}
and
\begin{equation}
\label{eq:fusion_prop_Upsilon}
\Upsilon(m_1; \theta+i \frac{\pi}{k}m_2) \Upsilon(m_2; \theta-i \frac{\pi}{k}m_1)=\Upsilon(m_1+m_2; \theta) \,.
\end{equation}
These properties can be used to obtain the Bethe--Yang equations for the bound states starting from those of the fundamental particles, as discussed below.

\paragraph{Left-left bound states.}
Consider a left particle with rapidity $\theta_1$ such that $\text{Im}(q_1(\theta_1))>0$. If this is the case, then in the infinite volume limit, it holds that $e^{ i q_1(\theta_1)} \to 0$. To avoid a zero on the r.h.s.\ of equation~\eqref{eq:left_particle_Bethe_Yang_eq} we require a pole in one of the S-matrix elements: such a pole can appear either in one of the dressing factors $\Phi(1,1;\theta_{1j})$ or in one of the dressing factors $\Phi(1,k-1;\theta_{1j})$. The first case should be interpreted as a particle of quantum number $m=2$ propagating in the $s$-channel  while the latter to a particle of quantum number $m=k-2$ propagating in the $t$-channel. Let us consider the first situation, for which a second particle must exist carrying a rapidity $\theta_2$ satisfying
\begin{equation}
\theta_1- \frac{i \pi}{k}=\theta_2+ \frac{i \pi}{k} \equiv \theta \implies \Phi(1,1; \theta_{12}) = \infty \,.
\end{equation}
Using~\eqref{eq:pminimum_theta1} and~\eqref{eq:rel_lim_p_and_E} it is immediate to check that the sum of the energies and momenta
of the two particles satisfy
\begin{equation}
\label{eq:fusionq1q1_E1E1}
\begin{aligned}
&q_1(\theta_1)+q_1(\theta_2)=q_2(\theta) \,,\\
&E_1(\theta_1)+E_1(\theta_2)=E_2(\theta) \,.
\end{aligned}
\end{equation}
The product of the Bethe--Yang equations for the two particles (neglecting the constant phases) becomes
\begin{equation}
\begin{split}
\label{eq:product_BY_eq_1_2}
    1 &= e^{\frac{2 \pi i L}{k} q_2(\theta)}  \prod_{j\neq 1,2}^{N_1} \Phi(2,1;\theta-\theta_j)\sigma(2,1;\theta-\theta_j)^{-2}\\
    & \times \prod_{j=1 }^{N_{\bar 1}} \Phi(2,k-1;\theta-\theta_j) \sigma(2,k-1;\theta-\theta_j)^{-2}
    \prod_{\alpha =1}^2 \prod_{j=1}^{N_y^{(\alpha)}} \Upsilon(2; \theta-\theta^{(\alpha)}_{j}),
\end{split}
\end{equation}
where to derive the expression above we used the fusion properties in~\eqref{eq:fusion_prop_dressing_phases} and~\eqref{eq:fusion_prop_Upsilon}.
If $\theta \in \mathbb{R}$ then $q_2(\theta)\in \mathbb{R}$ and $E_2(\theta)\ge0$; the r.h.s. of the equalities in~\eqref{eq:fusionq1q1_E1E1} can then be interpreted as the momentum and energy of a physical bound state with $m=2$. Since $q_2(\theta)\in \mathbb{R}$ then equation~\eqref{eq:product_BY_eq_1_2} is finite and the absence of poles in the  Bethe--Yang equation of the first particle implies the absence of poles in the equation of the second particle.
On the other hand, if $\theta_1+\theta_2 \notin \mathbb{R}$ then $q_2(\theta) \notin \mathbb{R}$ and~\eqref{eq:product_BY_eq_1_2} may still develop singularities in the limit $L \to \infty$. In this case, we need to keep going fusing.
The fusion procedure can continue until we reach a bound state composed of $k-1$ left particles and we obtain the Bethe--Yang equation for the right particles.

In general, we can fuse a number $Q$ of left particles (i.e. particles with quantum numbers $m_1=m_2=\dots=m_Q=1$) and construct bound states of any length $Q=1, \dots, k-1$.
The rapidities of the fused particle are 
\begin{equation}
\theta_j= \theta + \frac{i \pi}{k} (Q+1-2j) \hspace{3mm}, \hspace{3mm} j=1,\dots, Q \hspace{3mm}, \hspace{3mm} \theta \in \mathbb{R}
\end{equation}
and the associated energies and momenta satisfy
\begin{equation}
\sum_{j=1}^Q E_1(\theta_j)=E_Q(\theta)  \hspace{5mm} , \hspace{5mm}\sum_{j=1}^Q q_1(\theta_j)=q_Q(\theta) \,.
\end{equation}
Note that for $Q=k$ it holds that
\begin{equation}
\sum_{j=1}^k E_1(\theta_j)=0  \hspace{5mm} , \hspace{5mm}\sum_{j=1}^k q_1(\theta_j)=0 
\end{equation}
and the bound state obtained in this way is equivalent to the vacuum. Then the fusion procedure stops after $k-1$ iterations.

\paragraph{Right-right bound states.} 

One can apply the fusion procedure to the Bethe--Yang equations for right particles and obtain the equations for left particles. The procedure is slightly more convoluted than the one explained above since for right particles the bound states are obtained by fusing lower-weight states while the Bethe--Yang equations are written on a ground state composed of highest-weight states. In this case if we assume there is a right particle with momentum $q_{k-1}(\theta_1)$ having a positive imaginary part then the condition
\begin{equation}
\theta_1- \frac{i \pi}{k}=\theta_2+ \frac{i \pi}{k} \equiv \theta
\end{equation}
leading to 
\begin{equation}
\begin{aligned}
&q_{k-1}(\theta_1)+q_{k-1}(\theta_2)=q_{k-2}( \theta ) \,,\\
&E_{k-1}(\theta_1)+E_{k-1}(\theta_2)=E_{k-2}( \theta)\,,
\end{aligned}
\end{equation}
does not solve the vanishing problem of the Bethe--Yang equations.
Indeed, as outlined in~\cite{Frolov:2023lwd} $\Phi(k-1,k-1;\theta_{12})$ is singular but at the same time $\sigma(k-1, k-1; \theta_{12})^{-2}$ develops a zero of order two and the right-right S-matrix element $\Phi(k-1,k-1;\theta_{12}) \sigma(k-1, k-1; \theta_{12})^{-2}$ entering the right-particle equations produces a further zero. This is a consequence of the fact that the Bethe--Yang equations for the right particles were written in the $su(2)$ grading (the same problem happens in the pure Ramond-Ramond case~\cite{Frolov:2021bwp}). To cancel this zero one needs to obtain the right-particle bound state by fusing simultaneously two auxiliary roots. This zero of order two has to be cancelled by a pair of simultaneous poles in the auxiliary S-matrices:
\begin{equation}
\theta_1^{(1)}=\theta_1^{(2)}=\theta \implies \Upsilon(k-1; \theta_{1}-\theta_1^{(1)})=\Upsilon(k-1; \theta_{1}-\theta_1^{(2)})=\infty \,.
\end{equation}
Multiplying the four equations (the ones for the two physical particles and the ones for the pair of auxiliary roots) one obtains the Bethe equation for a bound state of $m=k-2$.
We should continue this procedure until we obtain a bound state with real momentum. If this bound state has been obtained by fusing $\bar{Q}$ right particles then the rapidities of the particles must satisfy
\begin{equation}
\theta_j= \theta + \frac{i \pi}{k} (\bar{Q}+1-2j) \hspace{3mm}, \hspace{3mm} j=1,\dots, \bar{Q} \hspace{3mm}, \hspace{3mm} \theta \in \mathbb{R} \,,
\end{equation}
and the associated auxiliary rapidities must be
\begin{equation}
\theta_j^{(\alpha)}=\theta + \frac{i \pi}{k} (\bar{Q}-2j) \hspace{3mm}, \hspace{3mm} j=1,\dots, \bar{Q}-1 \,.
\end{equation}
As for the Ramond-Ramond case~\cite{Frolov:2021bwp} the poles in the auxiliary S-matrices compensate the zeros in the physical S-matrices. The same bound state could have been obtained by fusing $k-\bar{Q}$ left particles. As expected from~\cite{Frolov:2023lwd} the fusion procedure closes with $k-1$ massive particles.

\paragraph{Absence of bound states in massless equations.}
Differently from massive particles, massless particles do not form bound states~\cite{Frolov:2023lwd}. At the level of the Bethe--Yang equations, this can be seen by noting that configurations of massless particles with complex rapidities always have non-positive energies and there are no allowed Bethe strings of massless particles.

\section{Derivation of TBA equations}
\label{app:derivation_TBA}

It is convenient to derive the TBA equations for a generic relativistic model and then specialise to the model under discussion. Consider a model with the following constraints for the root densities
\begin{equation}
\label{eq:rho_plus_rhobar_generic_rel_system}
\rho_k(\theta)+\bar{\rho}_k(\theta) = \frac{R}{2\pi} E_k+ \sum_{j} \rho_j * K^{jk}\,.
\end{equation}
The indices $k$ and $j$ label all the physical and auxiliary particles. The only difference is that $E_k\ne 0$ if $k$ is physical while $E_k= 0$ if $k$ is auxiliary.  The free energy of the model is given by
\begin{equation}
\mathcal{F}= \sum_k \int^{+\infty}_{-\infty} d \theta \Bigl(E_{k} \rho_k  - T s(\rho_k) - i T \gamma_k \rho_k \Bigl) \,,
\end{equation}
where again $E_k\ne0$ ($E_k=0$) for physical (auxiliary) particles.
The last term on the RHS of the equation above is associated with a chemical potential, where $\gamma_k$ are coefficients, one for each physical or auxiliary particle. Finally, $s$ is the density of entropy, given by
\begin{equation}
s(\rho) = \rho \log \Bigl(1 + \frac{\bar{\rho}}{\rho} \Bigl)+\bar{\rho} \log \Bigl(1 + \frac{\rho}{\bar{\rho}} \Bigl)\,.
\end{equation}

As usual, the thermodynamic equilibrium is reached when the free energy is minimal. 
Defining
\begin{equation}
\frac{\rho_k}{\bar{\rho}_k}= e^{i \gamma_k- \epsilon_k}
\end{equation}
and noting that 
\begin{equation}
\delta \rho_k+\delta \bar{\rho}_k = \sum_{j} \delta\rho_j * K^{jk}
\end{equation}
the variation of the entropy with respect to $\rho$ is given by\footnote{In performing the computation we note that the variation of the log's cancel by construction; indeed we have~$\rho\, \delta [\log ( 1+ {\bar{\rho}}/{\rho} )]+\bar{\rho}\, \delta[\log ( 1+ {\rho}/{\bar{\rho}} )]=0$.
}
\begin{equation}
\label{eq:entropy_variation}
\delta s(\rho_k)=\delta \rho_k \left( \epsilon_k - i \gamma_k \right)+\sum_{j} \bigl(\delta\rho_j * K^{jk}\bigl) \log \Bigl(1 + e^{i \gamma_k- \epsilon_k} \Bigl)
\end{equation}
The quantities $\epsilon_k$ are called pseudoenergies.
Plugging~\eqref{eq:entropy_variation} into the variation of the free energy we obtain
\begin{equation}
\delta \mathcal{F}= \sum_k \int^{+\infty}_{-\infty} d \theta \delta\rho_k \Bigl[E_{k}  - T \Bigl( \epsilon_k +\sum_{j}  \log \bigl(1 + e^{i \gamma_j- \epsilon_j} \bigl) * K^{jk}   \Bigl) \Bigl]
\end{equation}
from which we read off the equilibrium conditions
\begin{equation}
\label{eq:TBA_epsk}
\epsilon_k= \frac{E_k}{T} - \sum_{j} \log \bigl(1 + e^{i \gamma_j- \epsilon_j} \bigl) * K^{jk} \,.
\end{equation}
These equations, one for each particle type $k$, are known as the Thermodynamic Bethe Ansatz (TBA) equations.

Finally, using first~\eqref{eq:rho_plus_rhobar_generic_rel_system} to write $\bar{\rho}_k$ as functions of $\rho_k$ and then~\eqref{eq:TBA_epsk} we obtain
\begin{equation}
\label{eq:free_en_th_eq}
f(T)\equiv\frac{\mathcal{F}}{R}= -\frac{T}{2\pi} \sum_k \int^{+\infty}_{-\infty} d \theta \log \left( 1+ e^{i \gamma_k - \epsilon_k(\theta)} \right) E_k(\theta) \,.
\end{equation}
where we labelled by $f$ the free energy per unit of volume at the equilibrium.
It is common to introduce the Y functions
\begin{equation}
Y_k=e^{\epsilon_k} \iff Y_k=\frac{\bar{\rho}_k}{\rho_k} e^{i \gamma_k}
\end{equation}
and write the TBA equations~\eqref{eq:TBA_epsk} in terms of them as
\begin{equation}
\label{eq:TBA_Yk}
\log Y_k= \frac{E_k}{T} - \sum_{j}  \log \bigl(1 + \frac{e^{i \gamma_j}}{Y_j} \bigl) * K^{jk} \,.
\end{equation}

\paragraph{Identities for the model under consideration.}
In our model, we denote by $\mathcal{F}_m$, $\mathcal{F}_+$ and $\mathcal{F}_-$ the free energy associated with massive, chiral and antichiral particles, respectively. They are given by
\begin{equation}
\label{eq:variation_free_energy_massive}
\begin{split}
\mathcal{F}_m=&+ \sum_{P=1}^{k-1} \int_{-\infty}^{+\infty} d\theta \  \biggl[ \rho_P  E_P - T \Bigl[\rho_P \log \bigl( 1+\frac{\bar{\rho}_P}{\rho_P}\bigl)  +\bar{\rho}_P \log \bigl( 1+\frac{\rho_P}{\bar{\rho}_P} \bigl) \Bigl] \biggl]\\
&-\sum_{\alpha=1,2} T\sum_{\bbf= \uparrow \, \downarrow}  \int_{-\infty}^{+\infty} d\theta \  \biggl[\rho^{(\alpha)}_\bbf \log \bigl( 1+\frac{\bar{\rho}^{(\alpha)}_\bbf}{\rho^{(\alpha)}_\bbf}\bigl)  +\bar{\rho}^{(\alpha)}_\bbf \log \bigl( 1+\frac{\rho^{(\alpha)}_\bbf}{\bar{\rho}^{(\alpha)}_\bbf} \bigl) + i \gamma^{(\alpha)}_a \rho^{(\alpha)}_a\biggl] \,,
\end{split}
\end{equation}
\begin{equation}
\begin{split}
&\mathcal{F}_{\pm}=+ N_\circ \int_{-\infty}^{+\infty} d\theta \  \biggl[ \rho^{(\pm)}_0  E^{(\pm)}_0 - T \Bigl[\rho^{(\pm)}_0 \log \bigl( 1+\frac{\bar{\rho}^{(\pm)}_0}{\rho^{(\pm)}_0}\bigl)  +\bar{\rho}^{(\pm)}_0 \log \bigl( 1+\frac{\rho^{(\pm)}_0}{\bar{\rho}^{(\pm)}_0} \bigl) \Bigl] \biggl]\\
&\quad-\sum_{\alpha=1,2} T\sum_{\bbf= \uparrow \, \downarrow} \int_{-\infty}^{+\infty} d\theta \  \biggl[\rho^{(\pm, \alpha)}_\bbf \log \bigl( 1+ \frac{\rho^{(\pm, \alpha)}_{\bbf}}{\bar{\rho}^{(\pm, \alpha)}_{\bbf}} \bigl)  +\bar{\rho}_\bbf^{(\pm, \alpha)} \log \bigl( 1+\frac{\bar{\rho}^{(\pm, \alpha)}_{\bbf}}{\rho^{(\pm, \alpha)}_{\bbf}} \bigl) + i \gamma^{(\pm, \alpha)}_a \rho^{(\pm, \alpha)}_a \biggl] .
\end{split}
\end{equation}
We set the chemical potentials for the physical particles to zero and the ones for the massive and massless auxiliary roots to be $\gamma^{(\alpha)}_a$ and $\gamma^{(\pm, \alpha)}_a$.
Note that in the simplified equation ~\eqref{eq:constraint_rhoQ_densities2} the densities of roots and holes $\rho_\downarrow$ and $\bar{\rho}_\downarrow$ are exchanged compared to the generic case above. The same happens for $\rho^{(\pm, \alpha)}_\uparrow$ and $\bar{\rho}^{(\pm, \alpha)}_\uparrow$ in~\eqref{eq:massless_densities_simpl}. Then in the free energy we replace
\begin{equation}
\label{eq:replacement_rho_brho}
\begin{split}
&\rho_{\downarrow}^{(\alpha)}=-\bar{\rho}_{\downarrow}^{(\alpha)}+ \sum_{Q=1}^{k-1} \rho_Q \star K_{\downarrow}^{Qy}  \,,\\
&\rho^{(\pm, \alpha)}_{\uparrow}= -\bar{\rho}_{\uparrow}^{(\pm, \alpha)}+  N_\circ  \bigl[ \rho^{(\pm)}_0 \star s_2 \bigl]\,.
\end{split}
\end{equation}
Using that
\begin{equation}
\int^{+\infty}_{-\infty} dz \, K_{\downarrow}^{Qy}(z)= \frac{Q}{k}\,, \qquad \int^{+\infty}_{-\infty} dz \, s_2(z)= \frac{1}{2}
\end{equation}
we obtain
\begin{equation}
\begin{split}
\mathcal{F}_m=& \sum_{P=1}^{k-1} \int_{-\infty}^{+\infty} d\theta \  \biggl[ \rho_P  \left(E_P -i \frac{T P}{k} \sum_{\alpha=1,2} \gamma^{(\alpha)}_\downarrow \right) - T \Bigl[\rho_P \log \bigl( 1+\frac{\bar{\rho}_P}{\rho_P}\bigl)  +\bar{\rho}_P \log \bigl( 1+\frac{\rho_P}{\bar{\rho}_P} \bigl) \Bigl] \biggl]\\
&-\sum_{\alpha=1,2} T\sum_{\bbf= \uparrow \, \downarrow}  \int_{-\infty}^{+\infty} d\theta \  \biggl[\rho^{(\alpha)}_\bbf \log \bigl( 1+\frac{\bar{\rho}^{(\alpha)}_\bbf}{\rho^{(\alpha)}_\bbf}\bigl)  +\bar{\rho}^{(\alpha)}_\bbf \log \bigl( 1+\frac{\rho^{(\alpha)}_\bbf}{\bar{\rho}^{(\alpha)}_\bbf} \bigl) \biggl] \\
&- i T \sum_{\alpha=1,2} \int_{-\infty}^{+\infty} d\theta \ \left(\gamma^{(\alpha)}_\uparrow  \rho^{(\alpha)}_\uparrow -\gamma^{(\alpha)}_\downarrow  \bar{\rho}^{(\alpha)}_\downarrow \right) \,,
\end{split}
\end{equation}
\begin{equation}
\begin{split}
\mathcal{F}_{\pm}=&+ N_\circ \int_{-\infty}^{+\infty} d\theta \   \rho^{(\pm)}_0  \left(E^{(\pm)}_0 -\frac{i}{2} T \sum_{\alpha=1,2} \gamma^{(\pm, \alpha)}_\uparrow \right) \\
&- T N_\circ \int_{-\infty}^{+\infty} d\theta\left(\rho^{(\pm)}_0 \log \bigl( 1+\frac{\bar{\rho}^{(\pm)}_0}{\rho^{(\pm)}_0}\bigl)  +\bar{\rho}^{(\pm)}_0 \log \bigl( 1+\frac{\rho^{(\pm)}_0}{\bar{\rho}^{(\pm)}_0} \bigl) \right) \\
&-\sum_{\alpha=1,2} T\sum_{\bbf= \uparrow \, \downarrow} \int_{-\infty}^{+\infty} d\theta \  \biggl[\rho^{(\pm, \alpha)}_\bbf \log \bigl( 1+ \frac{\rho^{(\pm, \alpha)}_{\bbf}}{\bar{\rho}^{(\pm, \alpha)}_{\bbf}} \bigl)  +\bar{\rho}_\bbf^{(\pm, \alpha)} \log \bigl( 1+\frac{\bar{\rho}^{(\pm, \alpha)}_{\bbf}}{\rho^{(\pm, \alpha)}_{\bbf}} \bigl)\biggl]\\
&- i T \sum_{\alpha=1,2} \int^{+\infty}_{-\infty} d \theta \left( \gamma^{(\pm,\alpha)}_\downarrow \rho^{(\pm, \alpha)}_\downarrow - \gamma^{(\pm,\alpha)}_\uparrow \bar{\rho}^{(\pm, \alpha)}_\uparrow \right) \,.
\end{split}
\end{equation}
The replacement~\eqref{eq:replacement_rho_brho} translates into a shift of the particle energies. We can now obtain the TBA equations by requiring that at the equilibrium $\mathcal{F}$ is stationary under variations of the particle densities, with the further constraints that the equations~\eqref{eq:constraint_rhoQ_densities2} and~\eqref{eq:massless_densities_simpl} are satisfied.
Since massive, chiral and antichiral particles are decoupled we can perform these variations separately on $\mathcal{F}_m$, $\mathcal{F}_+$ and $\mathcal{F}_-$, finding the TBA equations in section~\ref{subsec:TBA_eq}.

\section{Ground state energy in the ultraviolet}
\label{sec:app_Rogers}

The ground state energy of a system at finite volume $L$ is connected to the free energy of the mirror theory at temperature $T=\frac{1}{L}$ through the relation~\cite{Zamolodchikov:1989cf}
\begin{equation}
\mathcal{E}=L f(\frac{1}{L})\,.
\end{equation}
For relativistic models the S-matrix is mapped into itself and after the mirror transformation equations~\eqref{eq:free_en_th_eq} and~\eqref{eq:TBA_epsk}
become
\begin{equation}
\label{eq:app_scaling_function_free_energy}
L \mathcal{E}= -\frac{1}{2\pi} \sum_k \int^{+\infty}_{-\infty} d \theta \ L E_k(\theta) \log \left( 1+ e^{- \epsilon_k(\theta)} \right)  \,.
\end{equation}
and
\begin{equation}
\label{eq:TBA_Yk_ap}
\epsilon_k= L E_k - \sum_{j} L_j * K^{jk} \,, \qquad L_j \equiv \log \bigl(1 + e^{-\epsilon_j} \bigl)\,,
\end{equation}
respectively.
In the latter equation $E_k=0$ if $k$ is an index associated with an auxiliary particle.
In this appendix we review how to determine the ground state energy in the UV limit $L m \ll1$ (here $m$ is a scale for the masses of the particles).

We consider a system in which all physical particles are massive with relativistic energies given by
\begin{equation}
E_k(\theta)= m_k \cosh{\theta}\,.
\end{equation}
Let us take the limit $m_k L \ll 1$. To compute the energy we need to split the rapidity line into three regions. The first region is 
\begin{equation}
E_k L \ll 1 \ \iff  -\log \frac{2}{ L m_k} \ll \theta \ll \log \frac{2}{L m_k} \,.
\end{equation}
In this region we can neglect the energy driving term in the TBA equations and the pseudo-energies reach a plateau where they are constant and need to satisfy
\begin{equation}
\label{eq:TBA_Yk_ap_plateau}
\begin{split}
&\epsilon^{(0)}_k= - \sum_{j}  \log \bigl(1 + e^{-\epsilon^{(0)}_j} \bigl) N_{jk} \,, \qquad N_{jk} \equiv -\int^{+\infty}_{-\infty} dt  K^{jk}(t)\,.
\end{split}
\end{equation}
Defining $Y^{(0)}_k \equiv e^{\epsilon^{(0)}_k}$, we have
\begin{equation}
\label{eq:Y_sysyem_constant_unsimplified}
Y^{(0)}_k= \prod_{j} \left( 1+ \frac{1}{Y^{(0)}_j} \right)^{N_{jk}} \,.
\end{equation}
Since in this region $E_k L \ll 1$ then there is a vanishing contribution to~\eqref{eq:app_scaling_function_free_energy}.

Next, consider the region 
\begin{equation}
E_k L \gg 1 \ \iff  \theta \ll -\log \frac{2}{L m_k} \ \cup \ \theta \gg \log \frac{2}{L m_k} \,.
\end{equation}
We want to find solutions for the auxiliary energies in this region.
In this region for physical particles the driving term $L E_k$ dominates and $\epsilon_k \to +\infty$.
If the particle is auxiliary the driving term is absent and if there is no auxiliary-auxiliary interaction, as in the models considered in the paper, then $\epsilon_k \to 0$. 
The reason is that we have
\begin{equation}
\epsilon^{\text{aux}}_k(\theta)= - \sum_{j} \int^{+\infty}_{-\infty} d t\log \bigl(1 + e^{-\epsilon^{\text{phys}}_j(t)} \bigl) * K^{jk}(t-\theta) \,.
\end{equation}
The only possibility for the $K^{jk}$ to do not be suppressed is that $t$ is in a region around $\theta$; since $\theta \gg 1$ then in this region $\epsilon^{\text{phys}}_j(t) \to +\infty$ and the integrand is then always suppressed. Then $\epsilon^{\text{aux}}_k(\theta) \to 0$ in this region. 
Defining $Y^{(\infty)}_k \equiv e^{\epsilon^{(\infty)}_k}$ we have
\begin{equation}
\label{eq:app_y_inifty}
\begin{split}
&Y^{(\infty)}_k= \infty \ \text{if} \ k \ \text{physical}\,, \qquad Y^{(\infty)}_k=1 \ \text{if} \ k \ \text{auxiliary}\,.
\end{split}
\end{equation}
Since in this region for physical particles $L_k \to 0$ there is a vanishing contribution to~\eqref{eq:app_scaling_function_free_energy}.

Finally, there is an interpolating region between these two domains, often called the kink region, where for $k$ physical $L_k$ interpolates between the plateau $L^{(0)}_k=\log \bigl(1 + e^{-\epsilon^{(0)}_j} \bigl)$ and $0$.
Let us focus on the tail of positive rapidity. Parameterising 
\begin{equation}
\theta=\log \frac{2}{L m_k}+ \tilde{\theta}
\end{equation}
then equation~\eqref{eq:TBA_Yk_ap} becomes
\begin{equation}
\label{eq:TBA_Yk_ap_kink}
\begin{split}
&\epsilon_k(\tilde{\theta})= e^{\tilde{\theta}}- \sum_{j} L_j * K^{jk} \qquad \text{if} \ k \ \text{phys}\,,\\
&\epsilon_k(\tilde{\theta})= 0 - \sum_{j} L_j * K^{jk}\qquad \ \  \text{if} \ k \ \text{aux}\,.
\end{split}
\end{equation}
A similar equation holds for negative rapidities, on the left tail of the kink.
Using that $L E_k=e^{\tilde{\theta}}$ for physical particles and $L E_k=0$ for the auxiliary ones we obtain the following expression for the ground state energy 
\begin{equation}
\label{eq:gr_state_kink_app}
\begin{split}
 L \mathcal{E}=& -\frac{1}{\pi} \sum_{k \in \text{phys}} \int^{+\infty}_{-\infty} d \tilde{\theta} e^{\tilde{\theta}} \log \left( 1+ e^{- \epsilon_k(\tilde{\theta})} \right)\\
 &-\frac{1}{\pi} \sum_{k \in \text{aux}} \int^{+\infty}_{-\infty} d \tilde{\theta} \ 0 \log \left( 1+ e^{- \epsilon_k(\tilde{\theta})} \right)\,.
 \end{split}
\end{equation}
Note that compared with~\eqref{eq:app_scaling_function_free_energy} we had to multiply by two due to the specular kink solution at negative rapidity.%
\footnote{From now on we will omit the tilde over the rapidity to lighten our notation.}
There is a trick to derive the ground state energy in this limit~\cite{Zamolodchikov:1989cf}. First we differentiate~\eqref{eq:TBA_Yk_ap_kink} wrt to $\theta$ and get 
\begin{equation}
\begin{split}
e^\theta= \epsilon'_k+ \sum_{j} \left[L'_{j} * K^{jk} \right](\theta) \qquad \text{if} \ k \ \text{phys}\,,\\
0= \epsilon'_k+ \sum_{j} \left[L'_{j} * K^{jk} \right](\theta) \qquad \text{if} \ k \ \text{aux}\,.
\end{split}
\end{equation}
Then we plug this formula into~\eqref{eq:gr_state_kink_app} and obtain
\begin{equation}
\label{eq:gr_state_kink_app2}
 L \mathcal{E}= -\frac{1}{\pi} \sum_{k}  \int^{+\infty}_{-\infty} d \theta \ \epsilon'_k(\theta) L_k(\theta) -\frac{1}{\pi} \sum_{k, j}  \int^{+\infty}_{-\infty} d \theta \  \left[L'_{j} * K^{jk} \right](\theta) L_k(\theta)  \,.
\end{equation}
Using now~\eqref{eq:TBA_Yk_ap_kink} the expression above becomes
\begin{equation}
\label{eq:gr_state_kink_app3}
\begin{split}
 L \mathcal{E}= &-\frac{1}{\pi} \sum_k  \int^{+\infty}_{-\infty} d \theta \ \epsilon'_k(\theta) L_k(\theta) \\
 &-\frac{1}{\pi} \sum_{j\in \text{phys}} \int^{+\infty}_{-\infty} d \theta L'_{j}(\theta) (e^\theta - \epsilon_j(\theta))-\frac{1}{\pi} \sum_{j\in \text{aux}} \int^{+\infty}_{-\infty} d \theta L'_{j}(\theta) (0 - \epsilon_j(\theta))\\
 = &-\frac{1}{\pi} \sum_k  \int^{+\infty}_{-\infty} d \theta \ \epsilon'_k(\theta) L_k(\theta) +\frac{1}{\pi} \sum_{j} \int^{+\infty}_{-\infty} d \theta L'_{j}(\theta) \epsilon_j(\theta) + \frac{1}{\pi} \sum_{j \in \text{phys}} \int^{+\infty}_{-\infty} d \theta L_{j}(\theta) e^\theta
 \end{split}
\end{equation}
where in the last line we performed an integration by part and got rid off the total derivative contribution. Noting that the last term in the expression above is the opposite of $L \mathcal{E}$ we end up with
\begin{equation}
2 L \mathcal{E}=-\frac{1}{\pi} \sum_k  \int^{+\infty}_{-\infty} d \theta \ \epsilon'_k(\theta) L_k(\theta) +\frac{1}{\pi} \sum_{k} \int^{+\infty}_{-\infty} d \theta L'_{k}(\theta) \epsilon_k(\theta)
\end{equation}
from which we can write
\begin{equation}
L \mathcal{E}=-\frac{1}{2\pi} \sum_k  \int^{\epsilon^{(\infty)}_{k}}_{\epsilon^{(0)}_{k}} d \epsilon_k \ \left( \log \left(1 + e^{- \epsilon_k} \right) + \frac{\epsilon_k \, e^{- \epsilon_k}}{1 + e^{- \epsilon_k}} \right) \,.
\end{equation}
The boundary of the integration $\epsilon^{(0)}_k$ are the solutions to~\eqref{eq:TBA_Yk_ap_plateau}, while $\epsilon^{(\infty)}_k$ are fixed by~\eqref{eq:app_y_inifty}.
After parameterising $e^{-\epsilon_k} = z_k$ the expression above can be written as
\begin{equation}
\begin{split}
L \mathcal{E}=&+\frac{1}{2\pi} \sum_k   \int^{1/Y^{(\infty)}_k}_{1/Y^{(0)}_k} d z \ \left( \frac{\log \left(1 + z \right)}{z} - \frac{\log z}{1 + z} \right) \,.
\end{split}
\end{equation}
The values of~$Y^{(\infty)}_k$ are determined on general considerations to be the ones in~\eqref{eq:app_y_inifty}; the expression above then becomes
\begin{equation}
\label{eq:app_gr_state_energy_UV}
\begin{split}
L \mathcal{E}=&+\frac{1}{2\pi} \sum_{j \in \text{phys}}   \int^{0}_{1/Y^{(0)}_j} d z \ \left( \frac{\log \left(1 + z \right)}{z} - \frac{\log z}{1 + z} \right)\\
&+\frac{1}{2\pi} \sum_{j \in \text{aux}}   \int^{1}_{1/Y_j^{(0)}} d z \ \left( \frac{\log \left(1 + z \right)}{z} - \frac{\log z}{1 + z} \right) \,.
\end{split}
\end{equation}
and is determined by the set of solutions to~\eqref{eq:Y_sysyem_constant_unsimplified}, which depends on the model under discussion.
Finding these solutions is non-trivial in general.

\paragraph{Remark on massless particles.}
For massless theories where the only non-vanishing interactions are of type chiral-chiral and antichiral-antichiral the analysis is almost identical to the one considered above. In this case we need to considered two decoupled systems of massless particles, which can be identified with the two tails of the kink solution previously discussed. The ground state energy is given by~\eqref{eq:app_gr_state_energy_UV} with Y functions obtained from~\eqref{eq:Y_sysyem_constant_unsimplified}.

\section{Simplification of the massive TBA equations}
\label{app:simplification_TBA}

We show how the TBA equations for massive particles
\begin{equation}
\label{eq:TBA_massive_phys_app}
\begin{split}
 \log Y_P= &+\frac{1}{T} E_P  -  \sum_{Q=1}^{k-1}   \log \bigl(1 +\frac{1}{Y_Q} \bigl) \star K_\Phi^{QP} \\
 &- \sum_{\alpha=1,2} \log \bigl(1 -\frac{e^{i \mu_\alpha}}{Y^{(\alpha)}_\uparrow}\bigl) \star K_{\uparrow}^{yP}- \sum_{\alpha=1,2} \log \bigl(1 -\frac{e^{-i \mu_\alpha}}{Y^{(\alpha)}_\downarrow}\bigl) \star K_{\downarrow}^{yP} \,,\\
 \log  Y^{(\alpha)}_a= &-\sum_{Q=1}^{k-1}   \log \bigl(1 +\frac{1}{Y_Q} \bigl) \star K_\bbf^{Q y} \,,
 \end{split}
\end{equation}
can be simplified. We start noting that the energies of massive-particle entering the TBA equations satisfy
\begin{equation}
E_P(\theta- \frac{i \pi}{k})+E_P(\theta+ \frac{i \pi}{k})= E_{P-1}(\theta)+E_{P+1}(\theta) \,.
\end{equation}
Similarly, the kernels describing the interaction between massive particles are related by
\begin{multline}
    \label{eq:Kernel_relations_with_epsilon1}
    K_\Phi^{QP}(\theta+\frac{i \pi}{k}- i \epsilon)+K_\Phi^{QP}(\theta-\frac{i \pi}{k}+ i \epsilon)-K_\Phi^{Q, P-1}(\theta)-K_\Phi^{Q, P+1}(\theta)= - \delta(\theta) I_{PQ} \,,
\end{multline}
\begin{multline}
 \label{eq:auxKernel_relations_with_epsilon1}
    K_{\bbf}^{yP}(\theta+\frac{i \pi}{k}-i \epsilon)+K_{\bbf}^{yP}(\theta-\frac{i \pi}{k}+i \epsilon)-K_{\bbf}^{y, P-1}(\theta)-K_{\bbf}^{y, P+1}(\theta)=\delta(\theta) \, I_{P \bbf} \,,
\end{multline}
where the incidence matrices $I_{P Q}$ and $I_{P \bbf}$ are defined as follows
\begin{equation}
\label{IPQ_IPbullet_definitions}
I_{PQ} \equiv \delta_{P-1, Q}+\delta_{P+1, Q} \quad\text{and} \quad I_{P \bbf} \equiv \delta_{P,1} \delta_{\bbf, \uparrow}+\delta_{P, k-1} \delta_{\bbf, \downarrow} \,.
\end{equation}
Acting with the Cauchy kernel $s_k$ (see~\eqref{eq:Cauchy_kernel_definition}) on~(\eqref{eq:Kernel_relations_with_epsilon1}, \eqref{eq:auxKernel_relations_with_epsilon1}) we obtain
\begin{equation}
\label{eq:relation_KPQ_Kbullet_convolutions}
\begin{split}
    &K_\Phi^{PQ}-\sum^{k-1}_{M=1} I_{PM} K_\Phi^{MQ}*s_k = - s_k I_{PQ} \,,\\
    &K_{\bbf}^{Py}-\sum^{k-1}_{M=1}I_{PM} K_{\bbf}^{M y}*s_k= s_k \, I_{P \bbf} \,.
    \end{split}
\end{equation}

Note that we cannot set $\epsilon=0$ in~\eqref{eq:Kernel_relations_with_epsilon1} and~\eqref{eq:auxKernel_relations_with_epsilon1} since when we translate the equations away from the real line by $\pm \frac{i \pi}{k}$ poles can overlap the integration line in general.
We can then sum the first line of~\eqref{eq:TBA_massive_phys_app} for two particles of quantum number $P$ and rapidities $\theta+\frac{i \pi}{k}$ and $\theta-\frac{i \pi}{k}$, and subtract from the result the same equation for two particles of rapidity $\theta$ and quantum numbers $P-1$ and $P+1$:
\begin{multline}
\label{eq:massive_simplified_app}
    \log \Bigl[ Y_P(\theta- \frac{i \pi}{k}+i \epsilon) Y_P(\theta+ \frac{i \pi}{k}-i \epsilon) \Bigl]=\log \Bigl[ \left(1+Y_{P+1}(\theta) \right) \left( 1+Y_{P-1}(\theta) \right) \Bigl] \\
    - \sum_{\alpha=1,2}  \delta_{P 1}   \log  \bigl(1 -\frac{e^{i \mu_\alpha}}{Y^{(\alpha)}_\uparrow}\bigl) -\delta_{P k-1}   \log  \bigl(1 -\frac{e^{-i \mu_\alpha}}{Y^{(\alpha)}_\downarrow}\bigl) \,.
\end{multline}
Exponentiating the result above we obtain the relation in~\eqref{eq:massive_simplified_exp}.

Let us now simplify the TBA equation for the auxiliary particles (which is the second row in~\eqref{eq:TBA_massive_phys_app}). To that purpose we use the crossing equations
\begin{equation}
\label{eq:appdistrib_cros_bb}
\langle Q \rangle_{\theta \pm \frac{i \pi}{2} \mp i \epsilon}+ \langle k-Q \rangle_{\theta \mp \frac{i \pi}{2} \pm i \epsilon}= \delta_{Q , \frac{k}{2}} \delta(\theta)\,.
\end{equation}
In order to make the simplification through crossing we need to shift $\theta$ by $\pm \frac{i \pi}{2}$ on the imaginary direction and pick up all poles overlapping the integration contour of the TBA equation when performing this shift. Suppose to start from the equation
\begin{equation}
\log Y^{(\alpha)}_\uparrow(\theta)= - \sum^{k-1}_{Q=1} \log \left(1+\frac{1}{Y_Q (z)} \right) * \langle Q \rangle_{z-\theta}\,.
\end{equation}
We now shift $\theta \to \theta + \frac{i \pi}{2} - i \epsilon$. As usual the parameter $\epsilon$ is a small positive regulator to avoid poles on the contour after the shift. The kernel $\langle Q \rangle_{z-\theta}$ has poles at
\begin{equation}
z=\theta \pm \frac{i \pi}{k}Q\,.
\end{equation}
When performing the shift the poles which originally are on the lower half of the complex plan cross the contour for all $Q<\frac{k}{2}$. Picking up the associated residues we obtain
\begin{equation}
\begin{split}
\log Y^{(\alpha)}_\uparrow(\theta+\frac{i \pi}{2} - i \epsilon)= &- \sum^{k-1}_{Q=1} \log \left(1+\frac{1}{Y_Q (z)} \right) * \langle Q \rangle_{z-\theta-\frac{i \pi}{2} + i \epsilon}\\
&- \sum_{Q< \frac{k}{2}} \log \left(1+\frac{1}{Y_Q(\theta+\frac{i \pi}{2} - \frac{i \pi}{k} Q - i \epsilon)} \right)\,.
\end{split}
\end{equation}
Repeting a similar shift for the other auxiliary Y functions we get
\begin{equation}
\label{eq:app_shifted_Yaux1}
\begin{split}
\log Y^{(\alpha)}_\uparrow(\theta+\frac{i \pi}{2} - i \epsilon)=& - \sum^{k-1}_{Q=1} \log \left(1+\frac{1}{Y_Q (z)} \right) * \langle Q \rangle_{z-\theta-\frac{i \pi}{2} + i \epsilon} \\
&- \sum_{Q< \frac{k}{2}} \log \left(1+\frac{1}{Y_Q(\theta+\frac{i \pi}{2} - \frac{i \pi}{k} Q - i \epsilon)} \right)\,,\\
\log Y^{(\alpha)}_\uparrow(\theta-\frac{i \pi}{2} + i \epsilon)=& - \sum^{k-1}_{Q=1} \log \left(1+\frac{1}{Y_Q (z)} \right) * \langle Q \rangle_{z-\theta+\frac{i \pi}{2} - i \epsilon}\\
&- \sum_{Q< \frac{k}{2}} \log \left(1+\frac{1}{Y_Q(\theta-\frac{i \pi}{2} + \frac{i \pi}{k} Q + i \epsilon)} \right)\,,\\
\end{split}
\end{equation}
and
\begin{equation}
\label{eq:app_shifted_Yaux2}
\begin{split}
\log Y^{(\alpha)}_\downarrow(\theta-\frac{i \pi}{2} + i \epsilon)=& - \sum^{k-1}_{Q=1} \log \left(1+\frac{1}{Y_Q (z)} \right) * \langle k-Q \rangle_{z-\theta+\frac{i \pi}{2} - i \epsilon}\\
&- \sum_{Q> \frac{k}{2}} \log \left(1+\frac{1}{Y_Q(\theta+\frac{i \pi}{2} - \frac{i \pi}{k} Q + i \epsilon)} \right)\,,\\
\log Y^{(\alpha)}_\downarrow(\theta+\frac{i \pi}{2} - i \epsilon)=& - \sum^{k-1}_{Q=1} \log \left(1+\frac{1}{Y_Q (z)} \right) * \langle k-Q \rangle_{z-\theta-\frac{i \pi}{2} + i \epsilon} \\
&- \sum_{Q> \frac{k}{2}} \log \left(1+\frac{1}{Y_Q(\theta-\frac{i \pi}{2} + \frac{i \pi}{k} Q - i \epsilon)} \right)\,.
\end{split}
\end{equation}
Summing now the first term in~\eqref{eq:app_shifted_Yaux1} with the first term in~\eqref{eq:app_shifted_Yaux2} (and similarly for the second terms of both equations) and using~\eqref{eq:appdistrib_cros_bb} we obtain
\begin{equation}
\begin{split}
&\log Y^{(\alpha)}_\uparrow(\theta+\frac{i \pi}{2})  Y^{(\beta)}_\downarrow(\theta-\frac{i \pi}{2})=- \sum_{Q=1}^{k-1} \log \left(1+\frac{1}{Y_Q(\theta+\frac{i \pi}{2} - \frac{i \pi}{k} Q)} \right)\,,\\
&\log Y^{(\alpha)}_\uparrow(\theta-\frac{i \pi}{2})  Y^{(\beta)}_\downarrow(\theta+\frac{i \pi}{2})=- \sum_{Q=1}^{k-1} \log \left(1+\frac{1}{Y_Q(\theta-\frac{i \pi}{2} + \frac{i \pi}{k} Q)} \right) \,.
\end{split}
\end{equation}
Exponentiating these equations we obtain the expressions in~\eqref{eq:aux_Y_simplified_new}.

\bibliographystyle{JHEP}
\bibliography{refs}

\end{document}